\documentclass[twocolumn]{aastex631}
\usepackage{amsmath}
\usepackage{graphicx}
\usepackage{subfigure}
\usepackage{rotating}
\usepackage{ragged2e}

\begin{document}

\title{Evaluating Mass Outflow Rate Estimators in FIRE-2 Simulations: \\ Towards a Self-Consistent Framework for Spectral Line Based Predictions}

\author[0000-0003-4166-2855]{Cody A. Carr}
 \affiliation{Center for Cosmology and Computational Astrophysics, Institute for Advanced Study in Physics \\ Zhejiang University, Hangzhou 310058,  China, codycarr24@gmail.com, renyuecen@zju.edu.cn}
\affiliation{Institute of Astronomy, School of Physics, Zhejiang University, Hangzhou 310058,  China}

\author[0000-0002-2838-9033]{Aaron Smith}
\affiliation{Department of Physics, The University of Texas at Dallas, Richardson, Texas 75080, USA}


\author[0000-0002-2499-9205]{Viraj Pandya}
\altaffiliation{NHFP Hubble Fellow.}
\affiliation{Columbia Astrophysics Laboratory, Columbia University, 550 West 120th Street, New York, NY 10027, USA}

\author[0000-0003-4073-3236]{Christopher C. Hayward}
\affiliation{Eureka Scientific, Inc., 2452 Delmer Street, Suite 100, Oakland, CA 94602, USA}
\affiliation{Kavli Institute for the Physics and Mathematics of the Universe (WPI), The University of Tokyo Institutes for Advanced Study, The University of Tokyo, Kashiwa, Chiba 277-8583, Japan}

 \author[0009-0002-9932-4461]{Mason Huberty}
\affiliation{Minnesota Institute for Astrophysics, School of
   Physics and Astronomy, University of Minnesota \\ 316 Church str
 SE, Minneapolis, MN 55455,USA}

\author[0000-0002-9136-8876]{Claudia Scarlata}
 \affiliation{Minnesota Institute for Astrophysics, School of
   Physics and Astronomy, University of Minnesota \\ 316 Church str
 SE, Minneapolis, MN 55455,USA}

\author[0000-0001-8531-9536]{Renyue Cen}
 \affiliation{Center for Cosmology and Computational Astrophysics, Institute for Advanced Study in Physics \\ Zhejiang University, Hangzhou 310058,  China, codycarr24@gmail.com, renyuecen@zju.edu.cn}
\affiliation{Institute of Astronomy, School of Physics, Zhejiang University, Hangzhou 310058,  China}



\begin{abstract}

Galactic outflows shape galaxy evolution, but their mass, energy, and momentum transfer remain uncertain. High-resolution spectroscopy can help, but systematic discrepancies hinder model interpretation.  In this study, we evaluate the performance of semi-analytical line transfer (SALT) and empirical partial covering models (PCMs) to recover the properties of outflows in the FIRE-2 simulation suite from synthetic Si II lines (1190Å, 1193Å, 1260Å, 1304Å, 1527Å).  When applicable, we assess each model’s ability to recover mass, energy, and momentum outflow rates, as well as radial density and velocity profiles, column densities, and flow geometries.  We find that the PCM underestimates column densities by 1.3 dex on average in the range $15 < \log{N}\ [\rm cm^{-2}] < 17$ with dispersion 1.3 dex.  We attribute this bias to instrumental smoothing.  Since the PCM underestimates column densities, it also underestimates flow rates, though its predictions are independent of radius, with a dispersion of 0.55 dex.  We detect no bias in the SALT estimates of the column density with dispersion 1.3 dex. When the velocity and density field obey power-laws, SALT can constrain the mass, momentum, and energy outflow rates to 0.36(0.63), 0.56(0.56), and 0.97(0.80) dex at $0.15(0.30)R_{\rm vir}$, respectively. However, certain profiles in FIRE-2 fall outside the SALT framework, where the model breaks down.  We find that SALT effectively tracks the flow geometry, capturing the temporal evolution of the photon escape fraction that is out of phase with the star formation rate, fully consistent with hydro-simulations.  We advocate for integral field unit spectroscopy to better constrain flow properties.  

\end{abstract}



\section{Introduction}\label{sec:intro}

Over the past several decades, the combined efforts of observational and theoretical research have led to profound insights into the inner workings of galaxy formation. Observations have revealed that massive, kiloparsec-scale, multi-phase outflows of gas, metals, and dust are continuously expelled from the interstellar medium (ISM) of star-forming galaxies across the universe \citep{Rupke2018, Veilleux2020}. Theoretical models and simulations suggest that these outflows are driven by various forms of feedback, including supernovae, radiation, stellar winds, relativistic jets, and cosmic rays \citep{Thompson2024}. These outflows are also essential for regulating star formation \citep{Hopkins2014}, limiting the growth of supermassive black holes \citep{DiMatteo2005}, and enriching the surrounding media with metals \citep{Oppenheimer2006, Tumlinson2017}. Consequently, outflows are now recognized as an integral component of the baryon cycle \citep{Donahue2022} and a primary factor influencing galaxy evolution \citep{Naab2017}.

While we understand the primary roles that galactic outflows play in galaxy formation, many of the underlying mechanisms remain a mystery. For instance, it is unclear what the dominant drivers of outflows are across different galaxy types, morphologies, and redshifts, how efficiently they operate, and at what rates they transfer mass, energy, and momentum to their surroundings. Understanding these rates is essential for comprehending the distribution of baryons in cosmic ecosystems (i.e., the baryon census, \citealt{Peroux2020}) and determining how feedback suppresses star formation \citep{Hopkins2014}. Feedback mechanisms are typically classified as either ejective or preventative. Ejective feedback drives outflows that can expel large quantities of mass from the gravitational potential wells of dark matter halos, inhibiting future cycles of star formation. In contrast, preventative feedback deposits significant amounts of energy into the ambient medium, increasing internal pressure and temperature to prevent cooling and lift cool clouds from the circumgalactic medium (CGM) \citep{Voit2017,Pandya2020,Carr2023(Chris),Pandya2023,Voit2024a,Voit2024b}.



Outflow efficiency is typically quantified using mass ($\eta_M$), momentum ($\eta_P$), and energy ($\eta_E$) loading factors (e.g., \citealt{Chisholm2017b}), which are defined as the respective outflow rates divided by the star formation rate (SFR). For example, $\eta_M$ serves as an important initial parameter in large-volume cosmological simulations, which often rely on sub-grid recipes to incorporate various feedback processes occurring on unresolved scales (e.g., \citealt{Springel2005, Vogelsberger2014}). However, different feedback implementations frequently produce similar large-scale predictions, such as the luminosity function, indicating a significant degree of degeneracy in these models \citep{Pandya2020,Mitchell2020}. As a result, $\eta_M$ is often treated as a poorly constrained parameter in cosmological simulations that must be constrained through observations.  \cite{Pandya2021} showed that “meso-scale” cosmological zoom-in simulations like FIRE-2 \citep{Hopkins2018} predict rather than prescribe these loading factors which makes it possible to study how the emergent properties of galactic winds vary with local and global conditions. In this paper we will ask whether it is possible to recover the mass, momentum and energy loading factors predicted by these simulations using mock observations.

Absorption line spectroscopy is a widely used and promising technique for constraining the properties of galactic outflows (e.g., \citealt{Rubin2014, Heckman2015, Zhu2015, Chisholm2016, Chisholm2017a, Carr2021a, Xu2022a}). Although spectral lines encode valuable information about the density and velocity of intervening gas, extracting this data is far from straight-forward. The challenge arises from the complex physics that governs the radiative processes responsible for the lines.  Currently, there is no universally accepted model interpretation.  On one hand, empirical methods, which aim to directly extract column densities from the spectral lines (e.g., \citealt{Rubin2014}), necessarily make assumptions to convert these measurements into flow rates, as seen in \cite{Xu2022a}.  On the other hand, physical models (e.g., \citealt{Scarlata2015,Krumholz2017,Carr2023}) that attempt to directly ascertain outflow rates often suffer from degeneracies, as highlighted by \cite{Gronke2015}.  As models increase in complexity, these degeneracies become more pronounced, necessitating higher-quality data to distinguish between different model predictions.  As such, analytical approaches for interpreting spectral lines focus on capturing only the essential underlying physics required to make sufficiently accurate predictions.  Determining which parameters are essential is challenging, however, but comparing simpler interpretive models with their more complex numerical counterparts can provide valuable insights. For instance, \cite{Carr2023} evaluated the biases in line transport models that arise from neglecting turbulent and thermal line broadening when solving the radiative transfer equation.

Recently, \cite{Carr2023(Chris)} highlighted a growing tension between the large mass loading factors ($\eta_M \sim 100$) typically used in cosmological simulations and the smaller values ($\eta_M < 10$) observed in UV studies of local low-mass galaxies ($10^7 < M_{\star}/\text{M}_{\odot} < 10^{11}$) (e.g., \citealt{Chisholm2017b, McQuinn2019}). Through straightforward analytical arguments, \cite{Carr2023(Chris)} demonstrated that the observed outflow rates in these low-mass galaxies are insufficient to eject enough mass to effectively suppress star formation, leading them to propose preventative feedback as an alternative explanation (see also \citealt{Pandya2020,Pandya2023,Voit2024a,Voit2024b}). These findings underscore the need to reconsider feedback mechanisms in simulations. Establishing the correct method by which feedback suppresses star formation requires accurate measurements of loading factors to properly inform simulations. Moreover, these values must be transferable to simulations, ensuring that the definitions of the relevant terms are consistent or at least well-defined.  Unfortunately, both of these criteria remain underdeveloped in the field.

Recently, \cite{Huberty2024} conducted a comparative analysis of semi-analytical line transfer (SALT) and partial covering model (PCM) estimates for outflow properties in objects from the COS LegAcy Spectroscopic SurveY (CLASSY, \citealt{Berg2022}) and found significant discrepancies. The models disagreed in their interpretation of column densities by up to two orders of magnitude and, in some cases, differed by an order of magnitude in the estimates of mass outflow rates. The large size of these discrepancies relative to the prescribed model uncertainties \citep{Xu2022a, Hu2023, Huberty2024} indicates that systematic uncertainties significantly hinder our ability to accurately extract information from spectral lines. We aim to address this problem by testing the ability of the various methods explored in \cite{Huberty2024} to recover outflow properties within the FIRE-2 simulation suite \citep{Hopkins2018}. Specifically, we will evaluate the performance of the SALT model developed by \cite{Carr2023} and the PCM version employed by \cite{Xu2022a} in predicting mass, momentum, and energy outflow rates. Additionally, we will assess their ability to predict other relevant quantities, such as velocity and density fields, column densities, and outflow geometries. This will be done using mock spectra generated from the FIRE-2 simulations in post-processing with the radiation transfer code COLT \citep{Smith2015}.

While the PCM has been previously tested against simulations (e.g., \citealt{delaCruz2021,Jennings2025}), this study, to our knowledge, represents the first direct comparison between semi-analytical and empirical methods using simulations as the ground truth. Our ultimate objective is to develop a self-consistent framework that links simulations with data interpretation techniques. This comparison will, at the very least, facilitate a direct comparison between simulation outputs and properties inferred from observational data, promoting more informed initial conditions and well-defined model predictions. Future spectrographs on 30-meter-class ground-based telescopes and 6-meter UV space telescopes will obtain spectra of galactic outflows with unprecedented resolution and signal-to-noise ratios. To leverage this data, we must reconcile the current systematic limitations that hinder the accuracy and robustness of their interpretation.

The remainder of this paper is organized as follows: Section 2 provides an overview of the FIRE-2 simulation suite, while Section 3 details the SALT and PCM models. In Section 4, we describe our process for generating mock spectra, compare these to spectra from the CLASSY survey, and outline our model fitting procedure. Section 5 tests the performance of each model, while Section 6 discusses their results, considering various caveats, including the impact of turbulent+thermal motion, and our interpretation of \cite{Huberty2024} in light of our findings. Finally, we present a summary of our conclusions in Section 7.

\begin{table*}[t]
\centering
\caption{Galaxy properties for the FIRE-2 simulation suite.  All values correspond to the range of values from the selected snapshots.  From left to right: (1) simulation, (2) redshift, (3) star formation rate, based on star particles $<20$ Myrs in age, (4) 3D stellar half-mass radius, (5) stellar mass, (6) virial mass, (7) virial radius, and (8) mass outflow rate of hydrogen.}
\begin{tabular}{cccccccccccc}
\hline \hline
Simulation & $z$ & $\rm SFR$ &$R_{1/2M_{\star}}$&$\log{ M_{\star}}$ & $\log{M_\text{vir}}$& $\log{R_\text{vir}}$&  $\dot{M}_\text{H}$ \\
&&$\text{M}_{\odot}\ \text{yr}^{-1}$&kpc&$\text{M}_{\odot}$&$\text{M}_{\odot}$&kpc&$\text{M}_{\odot}\ \text{yr}^{-1}$
&\\[.5 ex] \hline 
m10y &  0.0644--0.0922 &  0.000--0.00627 & 0.881--1.72 &  7.02 & 10.2 & 1.80 & 0.00295--0.255 \\
m10z &  0.00802--0.165 &   0.000--0.0842 &  1.91--6.11 & 7.49 & 10.5 & 1.92 & 0.00341--0.935 \\
m11a &    0.0735--0.203 &    0.000--0.172 &  1.71--5.39 & 8.02 & 10.6 & 1.92 &   0.0145--3.95 \\
m11b &  0.00480--0.114 & 0.00434--0.0319 &   2.23--4.51 & 8.02 & 10.6 & 1.95 &   0.0396--0.345 \\
m11c & 0.00802--0.0699 & 0.000279--0.534 &  2.69--8.70 & 8.92 & 11.1 & 2.13 &   0.0514--4.20 \\
m11q &  0.00480--0.174 &    0.0000--0.3609 &  2.48--8.82 &  8.59 & 11.2 & 2.12 &  0.00721--3.31 \\
m12f &  0.0608--0.0846 &      9.03--37.6 &  3.52--4.89 & 10.9 & 12.1 & 2.45 &      3.12--45.9 \\
m12i &     0.132--0.163 &      7.04--14.8 &    2.6--3.93 & 10.8 & 12.0 & 2.38 &     0.386--8.07 \\
m12m &  0.0178--0.0809 &      7.27--20.2 &  4.41--5.95 & 11.1 & 12.1 &  2.44 &     3.53--14.2 \\[1ex]
\hline
\end{tabular}
\label{tab:sims}
\end{table*} 



\section{Simulations} \label{sec:simulations}

All simulations used in this study refer to cosmological zoom-in simulations taken from the ``core'' FIRE-2 suite \citep{Hopkins2018,Wetzel2023}.  Simulation properties, aside from those obtained using COLT, were taken from the analyses conducted by \citet{Pandya2020} and \cite{Pandya2021}, which built on the works of \cite{Hopkins2014,Muratov2015,Muratov2017,Angles-Alcazar2017,Hafen2019,Hafen2020}. This includes the time-dependent positions and velocities of dark matter halo centers, which \citet{Pandya2020} measured using the Rockstar and consistent-trees codes \citep{Behroozi13a_rockstar,Behroozi13_trees}. Our spectral analysis focuses on outflows from various snapshots observed in a suite of 9 “core” FIRE-2 halos: 2 low-mass dwarfs with $M_{\rm{vir}} \sim 10^{10} \ \text{M}_{\odot}$ (m10y, m10z),  intermediate-mass dwarfs with $M_{\rm{vir}} \sim 10^{11} \ \text{M}_{\odot}$ (m11a, m11b, m11c, m11q), and 3 MW-mass halos with $M_{\rm{vir}} \sim 10^{12} \ \text{M}_{\odot}$ (m12i, m12f, m12m).  These halos were first presented in \cite{Wetzel2016,Garrison-Kimmel2017,Chan2018,Hopkins2018}. However, the comparison tests between outflow rate estimators are restricted to outflow episodes from m11c, the simulations with the largest mass still within the CLASSY sample, serving as a proof of concept.  Table~\ref{tab:sims} provides the range of simulation properties spanned by the selected outflow episodes from each simulation. 

\subsection{Flow Rates and Other Quantities}

In general, the FIRE-2 metal flow rates, $\dot{M}^{F}_{Z}$, are calculated on a particle-by-particle basis, within spherical shells, and scaled to the appropriate mass fraction, $Z$, of the relevant ion\footnote{Note that the metal ionic species mass fractions are computed in post processing with COLT. See Section~\ref{sec:mock_spectra}.}.  For a given shell, the flow rate is determined as follows:
\begin{eqnarray}
\dot{M}^{F}_{Z} = \sum_i \frac{Z_i m_iv_{r,i}}{\Delta L} \, ,
\label{eq:FIRE_MOR}
\end{eqnarray}
where the summation subscript $i$ denotes each selected particle within the shell, $Z_i$ is the mass fraction of the $i^\text{th}$ particle for the relevant ion, $m_i$ is the mass of the $i^\text{th}$ particle, $v_{r,i}$ is the radial velocity of the $i^\text{th}$ particle, and $\Delta L$ is the thickness of the shell \citep{Pandya2021}.  For simplicity, in this paper, outflows are defined by selecting particles with positive radial velocities ($v_r > 0$) and inflows by those with non-positive radial velocities ($v_r \leq 0$).  This implies that mass outflows rates ($\dot{M}^F_{\rm{out},Z}$) will be positive and mass inflow rates ($\dot{M}^F_{\rm{in},Z}$) will typically be negative.  We note that this velocity criterion allows for potential contributions to the flow rates from slower, cooler gas tracing turbulence \citep[see][for more discussion]{Pandya2021}.  Radially dependent flow rates are computed by dividing the virial radius, $R_{\rm{vir}}$, into 64 shells.  For time variation plots, flow rates are measured at radii of $0.15R_{\rm vir}$ and $0.3R_{\rm vir}$ in shells of thickness $0.1R_{\rm vir}$ centered at each radius.   

We use ray tracing to compute the metal column densities ($N^F_Z$) in the FIRE-2 simulations. To ensure that the derived values represent only the absorbing material affecting the observed spectrum, we consider only rays parallel to the line of sight that terminate on star particles. We have     
\begin{eqnarray}
    N_{Z}^F = \frac{1}{L_{\rm total}} \sum_j L_j\int_0^{R_\text{vir}} Z_j(l) n_j(l) \text{d}l \, ,
    \label{eq:FIRE_N}
\end{eqnarray}
where $n_j$ is the density along the $j^\text{th}$ ray, $L_j$ is the luminosity of the $j^\text{th}$ star particle, $L_{\rm total} = \sum_j L_j$ is the total stellar luminosity, and $\text{d}l$ represents the differential path length along the $j^\text{th}$ ray.  The sum runs over all star particles.  

Velocity and number density fields are calculated by summing the radial velocities and particle masses within the same sequence of shells used in the calculation of the flow rates. All other quantities that deviate from these definitions will be specified as needed.

\section{Models}

\label{sec:Models}


In our analysis, we focus on two key models for inferring galactic wind properties from observables: the semi-analytical line transfer (SALT) model studied by \cite{Carr2023}, and the version of the partial covering model (PCM) studied by \cite{Xu2022a} in application to the CLASSY spectra \citep{Berg2022}.  While SALT assumes an underlying physical description of an outflow, the PCM is largely empirically based.  Together, these models span the space of models used through out the literature to interpret spectral lines by representing the two extremes.  Our goals for this section are to introduce the assumptions, free parameters, and potential sources of uncertainty inherent to each model.  In addition, we will describe how to compute derived quantities such as the mass outflow rates.


\subsection{The Partial Covering Model (PCM)}

The PCM is one of the most widely used approaches for interpreting absorption lines, albeit many variations exist.  In the version used by \cite{Xu2022a}, the absorption profiles are assumed to be Gaussian and characterized by three parameters: the standard deviation, $\sigma$, amplitude, $A$, and central location, $\beta$.  As such, the first step in this model is to fit Gaussian-shaped line profiles to all absorption features.  This includes outflowing components, or blue-shifted absorption wells, and ISM components, or absorption wells centered at systemic velocity.  While this step introduces an unnecessary assumption --- namely, that the line profiles are Gaussian --- it also introduces several benefits which ameliorate subsequent calculations.  These include reducing the impact of noise, creating a straightforward method for interpolating values between adjacent spectral bins, and establishing a quantifiable method for identifying outflows.

After fitting the Gaussian profiles, \cite{Xu2022a} perform an F-test to determine whether the spectra show evidence for outflows.  To do this, they compare a pure ISM model to an identical model that includes outflowing components.  The ISM model consists of single Gaussian-shaped absorption wells, fixed at the centers of the lines under consideration. In contrast, the outflow model consists of the same model but with the addition of at least one free Gaussian component for each line.  If the F-value exceeds a specified confidence level, then an outflow is said to be detected.  All subsequent analysis regarding the outflows is performed using only the free Gaussian components.

The next step is to recover the covering fraction, $C_{f}^P$, and optical depth, $\tau^P$, at each frequency bin in the outflowing portion of the spectrum.  Since the PCM “views” each frequency bin in a spectrum, independently, it requires at least two lines from the same ion with different wavelength-oscillator strength products ($\lambda f$) to break the inherent degeneracy in absorption between $\tau^P$ and $C_{f}^P$ \citep{Arav2005}.  After fitting the Gaussian outflow model to both line profiles, $C_{f}^P$ and $\tau^P$ can be obtained from the corresponding absorption profiles using the following set of equations:
\begin{equation}
    \begin{cases}
\frac{I_{\rm 1}}{I_0} &= 1 - C_f(1-e^{-\tau^P})\\
\frac{I_{\rm 2}}{I _0} &= 1 - C_f(1-e^{-\omega_2\tau^P})\\
    &…\\
\frac{I_{\rm n}}{I _0} &= 1 - C_f(1-e^{-\omega_n\tau^P})\\
    \end{cases}       
\end{equation}
where $\omega_i$ is the ratio of the $\lambda f$ values with respect to a chosen reference line.  In the case of more than two lines, a minimization procedure can be performed to determine the values of $C_{f}^P$ and $\tau^P$ that best satisfy all the equations.

Subsequently, the column density, $N_{Z}^P$, can be obtained directly from a single line profile using the relation:
\begin{eqnarray}
    \tau^P(\nu) = \frac{\pi e^2}{m_e c} f N_{Z,\nu}^P \, ,
    \label{eq:tau_savage}
\end{eqnarray}
where $N_{Z,\nu}^P = \int \phi n(s) \text{d}s$ has units of column density per Hertz, and $\phi$ is the profile function governing the frequency ($\nu$) dependence of the interaction cross section.  Specifically, one isolates $N_{Z,\nu}^P$, then integrates the optical depth over the line profile. This can be expressed as:
\begin{eqnarray}
    N_{Z}^P = \int \frac{m_e c}{\pi e^2} f \tau^P d\nu^{\prime} = \frac{m_e c}{\pi e^2} f \int \ln{\left(\frac{I_0}{I}\right)} d\nu^{\prime} \, , \label{eq:AOD}
\end{eqnarray}
where the integration is performed over the observed velocity range of the line profile, represented by $I/I_0$.  

With the column density in hand, \cite{Xu2022a} then compute the mass outflow rate, $\dot{M}_{\rm{out},Z}^P$, as 
\begin{eqnarray}
    \dot{M}_{\rm{out},Z}^P = 4\pi m_{Z} R_\text{SF} \int N_{Z,v^{\prime}}^P v^{\prime} \text{d}v^{\prime} \, , \label{eq:PCM_MOR}
\end{eqnarray}
where the integral is again taken over the observed velocity range, $N_{Z,v^{\prime}}^P$ now has units of column density per unit velocity, $m_{Z}$ is the mass of the relevant ion, and $R_{\rm{SF}}$ is the radial scale of the outflow.  In deriving Equation~\ref{eq:PCM_MOR}, \cite{Xu2022a} assume a constant outflow rate, a constant velocity field, a $r^{-2}$ density field, and a spherical outflow geometry.  While these assumptions are not strictly necessary, they were reasonably justified for the CLASSY galaxies.  We choose to adopt the same assumptions and save our comments on the associated uncertainty for later on.  It follows that the momentum outflow rate, $\dot{P}$, and energy outflow rate, $\dot{E}$, can be computed as
\begin{eqnarray}
     \dot{P}_{\rm{out},Z}^P = 4\pi m_{\rm ion} R_\text{SF} \int N_{Z,v^{\prime}}^P {v^{\prime}}^2 \text{d}v^{\prime}
     \label{eq:PCM_POR}
\end{eqnarray}
and 
\begin{eqnarray}
     \dot{E}_{\rm{out},Z}^P = 2\pi m_{\rm ion} R_\text{SF} \int N_{Z,v^{\prime}}^P {v^{\prime}}^3 \text{d}v^{\prime},
     \label{eq:PCM_EOR}
\end{eqnarray}
respectively. We list the free parameters of the PCM in Table~\ref{t3}.

\begin{table}
\caption{PCM Free Parameters, Descriptions, and Ranges.}
\centering
\resizebox{\columnwidth}{!}{$\begin{tabular}{|c|c|c|}
\hline\hline
Parameter & Description & Range \\[.5 ex]
\hline
\multicolumn{3}{|l|}{Outflow}\\
\hline
$\sigma_{\rm Out,
line}^P$ & Gaussian standard deviation & $[0,\infty)$ \\
$A_{\rm Out,line}^P$ & Gaussian height & $[0,1]$ \\
$\beta_{\rm Out,line}^P$ & Gaussian location & $(-\infty,\infty)$ \\
$\tau^P$ & optical depth & $[0,\infty)$ \\
$C_f^P$ & covering fraction & $[0,1]$ \\
\hline
\multicolumn{3}{|l|}{ISM}\\
\hline
$\sigma_{\rm ISM,
line}^P$ & Gaussian standard deviation & $[0,\infty)$ \\
$A_{\rm ISM,line}^P$ & Gaussian height & $[0,1]$ \\
\hline
\end{tabular} $}
\label{t3}
\end{table} 

In theory, Equation~\ref{eq:AOD} should provide an exact measurement of the column density of material in front of a point source, regardless of the velocity structure of the intervening material \citep{Spitzer1978}. This capability is arguably the greatest strength of the PCM. However, in practice, the line profile, $I/I_0$, may be smeared by instrumental smoothing.  Given the non-linearity of the integrand, Equation~\ref{eq:AOD} will only equal the true column density of the outflow under two conditions: 1) the velocity structure of the line profile is fully resolved, or 2) the outflow is sufficiently optically thin across all frequencies, such that $\ln I_0/I \sim 1-I/I_0$ everywhere \citep{Jenkins1996}.  When these conditions are not met, we say Equation~\ref{eq:AOD} measures an apparent column density (see \citealt{Savage1991}).  

A similar source of error can occur if the assumed Gaussian structure for the absorption profiles does not fit the data well.  For instance, if the line profile becomes saturated at more than one point.  See \cite{Huberty2024} for examples of non-Gaussian spectra observed in CLASSY \citep{Berg2022}.  

\cite{Huberty2024} found that the apparent column density underestimated the true column density when applying Equation~\ref{eq:AOD} to mock spectra of the Si II 1190\AA, 1193\AA\ doublet derived from simulations of idealized outflows \citep{Carr2023}.  The discrepancy increased at higher column densities and exceeded an order of magnitude for column densities $\log N_{\rm Si^{+}}\ [\rm cm^{-2}] > 16$ at a spectral resolution of $20\rm \ km\ s^{-1}$.  The simulations assumed power-law structures for the density and velocity fields of the outflows, as well as a radially dependent microturbulence that scales with the velocity field as $0.1\,v(r)$.  Given that the discrepancy between apparent and true column densities depends on the shape of the line profile, it is uncertain how severe this issue will be when Equation~\ref{eq:AOD} is applied to real data, where deviations from the idealized models are possible.  Our tests on the more complex outflows within the FIRE-2 simulations should help clarify this uncertainty.

An additional source of uncertainty arises from the fact that the PCM does not account for variations in the density field along different lines of sight other than the binary scenario afforded by a partial covering of the source.  According to Jensen's Inequality,
\begin{eqnarray}
    e^{-\langle\tau\rangle} \leq \langle e^{-\tau} \rangle \, ,
    \label{eq:Jensen's_Inequality}
\end{eqnarray}
the absorption occurring over multiple lines of sight (right side of the equation) may be less than what is predicted by an average optical depth, $\langle\tau\rangle$ 
(left side of the equation).  Since the PCM typically corresponds to the left hand side of Equation~\ref{eq:Jensen's_Inequality}, it may underestimate the column density (see \citealt{Jennings2025} for more discussion).

Lastly, the failure to account for resonant scattered emission in addition to absorption introduces additional uncertainty.  For instance, blue emission in-filling \citep{Prochaska2011,Zhu2015} can act to partially fill in the absorption feature and lead to an underestimate of the column density by the PCM.  Furthermore, since absorption only acts as a probe of gas positioned along the line of sight, it is not a strong probe of the full spatial extent of outflows, which are capable of extending tens of kiloparsecs from the galactic center \citep{Shaban2022}.  As such, a failure to capture the full extent of the outflow may result in uncertainties on the order of at least $4\pi$ \citep{Chisholm2016,Carr2018}.  

\subsection{The Semi-Analytical Line Transfer (SALT) Model}


SALT is a radiation transport model designed to predict the emission and absorption features observed in the spectra of galactic outflows \citep{Scuderi1992,Scarlata2015,Carr2018,Carr2021a,Carr2023}. In contrast to the PCM, SALT is a forward model and assumes an underlying outflow structure with a well defined geometry, velocity structure, density field, etc.  When paired with a Monte Carlo fitting procedure, one can constrain the SALT parameter space to infer the properties of outflows \citep{Carr2021a,Huberty2024}.  Our analysis will focus on the most recent version of the model presented in \cite{Carr2023}.  As the model is more complicated than the PCM, we provide only a summary of the relevant features here and refer the reader to the literature for the details.  



\begin{figure}
    \centering
    \includegraphics[scale=0.48]{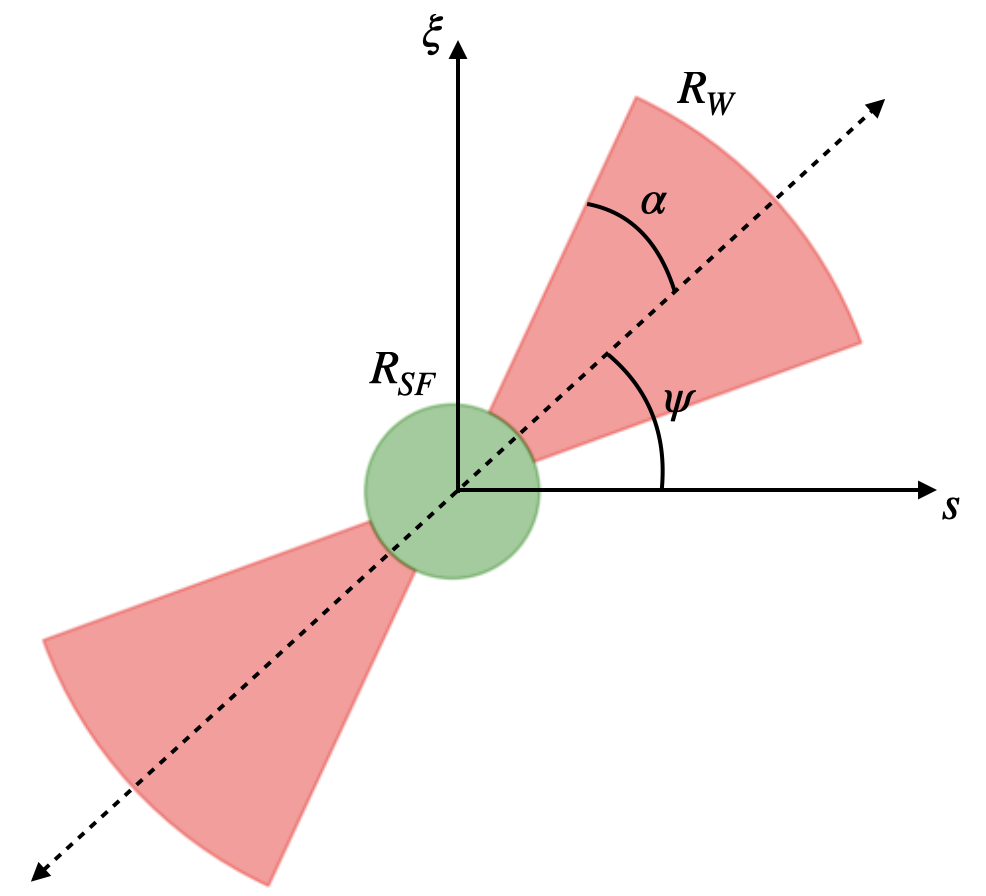}
    \caption{A cross-sectional view of the idealized outflow underlying the SALT calculations. The model consists of a spherical source ($r=R_{\rm SF}$) of isotropic radiation surrounded by a bi-conical outflow of half-opening angle, $\alpha$, and orientation angle, $\psi$, which extends to a terminal radius, $R_{\rm W}$.  The $s$ and $\xi$-axes point in directions parallel and perpendicular to the observer, respectively, and are measured in units of distance normalized by $R_\text{SF}$.}
    \label{fig:bi-cone}
\end{figure}

The SALT calculations are performed on an idealized model of a galactic outflow.  The base model features a spherical source of isotropic radiation with a radius of $R_{\rm SF}$, surrounded by a bi-conical outflow that extends to a terminal radius of $R_{\rm W}$.  Physically, the source is meant to embody the star forming regions of a galaxy from which an outflow is expelled.  In this work, we set $R_{\rm SF}$ equal to the 3D stellar half-mass radius.  The bi-cone is characterized by an opening angle, $\alpha$, and an orientation angle, $\psi$, subtended by the line of sight and axis of the bi-cone.  A diagram of the underlying model is provided in Figure~\ref{fig:bi-cone}.  

The outflow is further characterized by a power law density field of the form,
\begin{eqnarray}
n(r) = n_0\left(\frac{R_\text{SF}}{r}\right)^{\delta} \, ,
\end{eqnarray}
where $n_0$ is the density at $R_\text{SF}$ and $\delta$ is the power-law index.  Likewise, the velocity field, $v$, of the outflow assumes a power law of the form,
\begin{equation}
\begin{aligned}
v(r) &= v_0\left(\frac{r}{R_{\text{SF}}}\right)^{\gamma} &&\text{for}\ r < R_\text{W} \\[1em]
v &= v_{\infty}  &&\text{for} \ r \geq R_\text{W} \, , \\
\end{aligned}
\label{Velocity_Equation}
\end{equation} 
\noindent where $v_0$ is the launch velocity at $R_\text{SF}$, $v_{\infty}$ is the terminal velocity at $R_W$, and $\gamma$ is the power-law index.  The porosity of the outflow is described by the parameter, $f_c$, which represents the fraction of a bi-conical shell covered by material.  $f_c$ is taken to be constant with radius (see \citealt{Carr2018,Carr2021a}).

SALT computes the absorption and re-emission (resonant+fluorescent) of photons by the outflow for bound-bound transitions assuming the Sobolev approximation \citep{Lamers1999}.  Absorption is controlled by the free parameter, $\tau^S$, which is related to the optical depth, $\tau_0$, through the relation, $\tau^S = \tau_0/f_{ul}\lambda_{ul}$, where $f_{ul}$ and $\lambda_{ul}$ are the oscillator strength and wavelength from the relevant transitions, respectively.  $n_0$ can be recovered from $\tau$ through the relation,
\begin{eqnarray}
    \tau^S = \frac{\pi e^2}{m_ec}n_0\frac{R_\text{SF}}{v_0} \, . 
\end{eqnarray}

Unlike the PCM, SALT can constrain $\tau^S$ from a single line profile.  Because SALT assumes a structure for the outflow, it “views” the line profile as a function of frequency, and can constrain $\tau^S$ from the gradient of the line profile.  Moreover, by extrapolating the assumed analytical forms, constrained to the optically thin portions of a spectrum, SALT can constrain the optically thick or saturated regions of a line profile.   

SALT can account for observational effects such as a limiting circular observing aperture with projected radius, $R_{\rm AP}$, on the plane of the sky.  $v_{\rm ap}$ represents the value of the velocity field at $R_{\rm AP}$ and designates the free parameter in SALT controlling the aperture.  A failure of the aperture to capture the full extent of an outflow will result in reduced emission (see \citealt{Scarlata2015}).  SALT can account for dust extinction in the CGM assuming a fixed dust-to-gas ratio.  This parameter is controlled by $\kappa$ which is related to the dust opacity (see \citealt{Carr2021a}).


SALT does not perform radiation transfer modeling at radii less than $R_{\rm SF}$.  Therefore, to account for attenuation by the ISM, a Gaussian shaped absorption well, centered on each absorption line, is removed from the background flux (i.e., the continuum) before it is propagated through the outflow (see \citealt{Carr2023} for details).  In this work, we assume that all photons absorbed at resonance in the ISM reappear in fluorescence --- that is, we include equal, but inverted, Gaussian profiles at each corresponding fluorescent emission line.  Note that, in the absence of dust, coherent scattering will act to fill in ISM absorption.  In other words, ISM attenuation will cancel itself out.  In the case of fluorescent scattering, however, only a fraction of the ISM absorption will be filled in, with the exact amount depending on the optical depth and Einstein coefficients (see \citealt{Scarlata2015} for an explanation).  Thus, our ISM model is reasonable for this analysis since all lines in consideration have fluorescent components and the simulations naturally have high column densities and low dust-to-gas ratios.  The ISM adds two parameters for every absorption line to the SALT model: the amplitude, $A^S$, and standard deviation, $\sigma^S$.

It is important to note that while we have removed all re-emission from scattering in the ISM to fluorescent wavelengths, ISM absorption can still be filled in by resonant re-emission occurring in the outflow \citep{Prochaska2011,Scarlata2015}.  SALT should be able to handle this scenario, as it models both the absorption and re-emission of photons by the outflow.  This is in contrast to the PCM which should underestimate the amount of ISM absorption when this is the case.   

In total, the outflow component of the SALT model has 10 free parameters and the ISM component adds two additional parameters for every absorption line considered.  The SALT parameter names and descriptions are listed in Table~\ref{tab:salt_free_parameters}. 

In the SALT formalism, the metal mass outflow rate, $\dot{M}_{\rm{out},Z}^S$, can be computed directly from the free parameters as,
\begin{eqnarray}
\dot{M}_{\rm{out},Z}^S(r) &=& 4\pi(1-\cos{\alpha)}R_\text{SF}^2 \nonumber\\
&& \times m_{\rm ion}n_{0,\rm ion} \left( \frac{r}{R_{\rm SF}} \right)^{2- \delta} v\, ,
\label{eq:SALT_MOR}
\end{eqnarray}
where $m_{\rm ion}$ and $n_{0,\rm ion }$ are the mass and number density at $R_\text{SF}$ of the relevant ion, respectively.  Furthermore, the momentum outflow rate, $\dot{P}_{\rm{out},Z}^S$ can be computed as 
\begin{eqnarray}
 \dot{P}_{\rm{out},Z}^S(r)=\dot{M}_{\rm{out},Z}^S v
 \label{eq:SALT_POR}
\end{eqnarray} 
and the energy outflow rate, $\dot{E}_{\rm{out},Z}^S$, as
\begin{eqnarray}
    \dot{E}_{\rm{out},Z}^S(r)=\frac{1}{2}\dot{M}_{\rm{out},Z}^Sv^2.
    \label{eq:SALT_EOR}
\end{eqnarray}
Note that we are neglecting the contribution of the wind enthalpy (thermal energy + capacity to expand via $PdV$ work) in our definition of $\dot{E}^S$.  Similarly, we are neglecting the thermal pressure contribution to the momentum flux in $\dot{P}^S$.  See equations 7-8 of \cite{Pandya2021} and associated text for more information.    

The column density can be computed directly from the free parameter space as well.  However, because SALT assumes an extended source, care must be taken to account for variations in the density field along different lines of sight.  To do this, the column density is averaged over different paths after randomly sampling over the surface of the source.  The column density, $N_{Z,l\xi}^S$, along an arbitrary sight line can be computed as
\begin{eqnarray}
N_{Z,l\xi}^S = \int_{L}^{U}\Phi_{\rm cone}n_{0,\rm ion}\left(\frac{l^2+{s^{\prime}}^2+\xi^2}{R_\text{SF}^2}\right)^{-\delta/2}\text{d}s^{\prime} \, ,
\label{eq:Nse}
\end{eqnarray}
where
\begin{eqnarray}
U = (R_W^2-l^2-\xi^2)^{1/2} \, ,
\end{eqnarray}
\begin{eqnarray}
L = (R_\text{SF}^2-l^2-\xi^2)^{1/2} \, ,
\end{eqnarray}
and
\begin{gather} 
\Phi_{\rm cone} \equiv 
\begin{cases}
1 &\ \rm{if} \  (s^{\prime},l,\xi) \in \text{bi-cone} \\[1em]
0 & \ \rm{otherwise}. \\[1em]
\end{cases}
\end{gather}
The indicator function, $\Phi_{\rm cone}$, accounts for the bi-conical geometry of the outflow and the $l$-coordinate is perpendicular to the $s\xi$-plane (see Figure~\ref{fig:bi-cone}).  A more detailed description with a diagram is provided in the Appendix of \cite{Huberty2024}.  To achieve the final column density, $N_{Z,}^S$, responsible for the line profile observed in SALT, we then average over multiple lines of sight:
\begin{equation}
    N_{Z}^S = I^{-1} \sum_i^I N^S_{Z,l(i)\xi(i)} \, ,
\end{equation}
where the sum occurs over $I$ lines of sight drawn uniformly over the surface of the source.  We find that the values of $N_{Z}^S$ start to converge to roughly two significant figures by $I = 10^4$ in agreement with the Poisson error estimate of $I^{-1/2}$.

\begin{table}
\centering
\caption{SALT Model Parameters}
\resizebox{\columnwidth}{!}{$\begin{tabular}{cccc}
\hline\hline
Parameter & Description & Prior&ICs \\[.5 ex]
\hline
\multicolumn{3}{l}{Outflow}\\
\hline
$\alpha$ & half-opening angle [rad] & $[0,\pi/2]$& $[0,\pi/2]$ \\
$\psi$ & orientation angle [rad] & $[0,\pi/2]$& $[0,\pi/2]$ \\
$v_0$ & launch velocity [$\mathrm{km \ s^{-1}}$] & $[0,150]$ & $[0,150]$\\
$v_{\infty}$ & terminal velocity [$\mathrm{km \ s^{-1}}$] & $[150,1200], \ > v_0$ & $[200,800]$ \\
$v_{ap}$ & velocity at $R_{AP}$ [$\mathrm{km \ s^{-1}}$] & $[200,1200]$ & $[200,800]$\\
$\gamma$ & velocity field power law index & $[0.5,2]$ &$[0.5,2]$\\
$\delta$ & density field power law index & $0.5<\delta-\gamma<3.5$& $[-1.5,1.5]$ \\
$\tau$ & optical depth divided by $f_{ul}\lambda_{ul}$ [$\text{\AA}^{-1}$] & $-3<\log{\tau}<1 $& $[-2,1]$\\
$f_c$ & covering fraction & $[0,1]$ & $[0,1]$\\
$\kappa$ & dust opacity multiplied by $R_\text{SF}n_{0,\text{dust}}$ & $[0,50]$ & $[0,50]$ \\
\hline
\multicolumn{3}{l}{ISM}\\
\hline
$A_{\rm line}$ & ISM Gaussian absorption depth & $[0,1]$ & $[0,1]$ \\
$\sigma_{\rm line}$ & ISM Gaussian standard deviation [$\mathrm{km \ s^{-1}}$] & $[0,300]$ & $[0,300]$ \\
\hline
\end{tabular} $}
\begin{minipage}{\columnwidth}
\justifying
\small
\noindent From left to right: parameter (1) name (2) description (3) Prior (4) Initial Conditions.  See \cite{Carr2023} for additional definitions. Dashed lines refer to constrained quantities.  
\end{minipage}
\label{tab:salt_free_parameters}
\end{table}
 
We now consider various sources of uncertainty for SALT.  As previously stated, SALT relies on the Sobolev approximation \citep{Lamers1999} to solve the radiation transfer equation \citep{Chandrasekhar1960}. As such, SALT does not account for turbulent or microscopic motion in the outflow. \cite{Carr2023} showed that the presence of turbulent/thermal line broadening can lead to biases in the recovery of some parameters (e.g., $v_0$), which in turn can lead to biases in derived quantities such as the outflow rate and column density.  The existence and magnitude of such a bias will depend on the Doppler parameter of the gas, as well as the relative strength of the velocity field \citep{Carr2023}.  Tests against the FIRE-2 simulations will allow us to better gauge how large this uncertainty is in real galaxy conditions.  While SALT does consider variations in column density along different lines of sight, it does not account for variations in surface brightness \citep{Marconcini2023}.  This may lead to an incorrect weighting of different sightlines as expressed in Equation~\ref{eq:FIRE_N}.  As previously stated, SALT does not perform radiation transfer modeling in the ISM. While we can essentially circumvent this problem by removing an ISM absorption feature from the continuum before its propagation through the outflow, it is unclear how such a feature will affect parameter recovery.  Finally, deviations from the idealized constructs assumed by SALT (e.g., non bi-conical outflow geometries), are an obvious source of uncertainty.  

\section{Synthetic Spectra}
\label{sec:mock_spectra}


In this section, we outline our procedure for generating mock spectra of the FIRE-2 simulations.  Our goal is to produce realistic snippets, or spectral cutouts, of any desired line profile, along any line of sight, within any simulation. Since the FIRE-2 simulations do not track individual ions during runtime, we must handle this aspect in post-processing.  To achieve this, we use the latest version of the Cosmic Ly$\alpha$ Transfer code \citep[COLT;][]{Smith2015,Smith2019,Smith2022}, to account for the different metal ionization states for spatially-resolved resonance line opacities. COLT is capable of handling a variety of grid types for hydrodynamic codes, including Voronoi tessellations constructed from a set of mesh-generating points, which we adopt as a natural representation of gas properties provided by the meshless finite mass method employed by GIZMO \citep{Hopkins2015}, making it an ideal tool for the FIRE-2 simulations.


We briefly describe the photoionization equilibrium calculations used in our study. Ionizing photons are primarily emitted by young, massive stars, with stellar populations represented by star particles in the simulation. These sources are characterized by spectral energy distributions (SEDs) adopted from the Binary Population and Stellar Synthesis (BPASS) models (version 2.2.1; \citealt{Stanway2018}).  In particular, we use the \verb"bin-imf135_100" model with binary stars that assumes a Salpeter initial mass function (IMF; \citealt{Salpeter1955}) with lower and upper mass cutoffs of 0.1 and 100 $\text{M}_{\odot}$, respectively, and slope $\alpha_s = -2.35$. To ensure well-sampled stellar emission in the Monte Carlo radiative transfer (MCRT) calculations, we apply a luminosity boosting technique, adding statistical power to less luminous sources by biasing the photon packet distribution. Ionizing photons are sampled into energy bins determined by the ionization thresholds of all included atomic species (H, He, C, N, O, Ne, Mg, Si, S, Fe), ensuring sufficient spectral resolution to accurately track ionization processes.

Beyond stellar photoionization, COLT also accounts for recombinations, collisional ionizations, charge exchange reactions, and a meta-galactic UV background with self-shielding \citep{Faucher-Giguere:2009aa, Rahmati:2013aa}. Photon packets are launched isotropically from star particles and accurately propagated through the ISM using ray-tracing techniques, with dust and gas absorption modeled continuously. As there is no on-the-fly dust model, we adopt a global dust-to-metal ratio of 0.4, with the dust opacity, albedo, and anisotropic scattering parameters taken from the Milky Way dust model of \citet{Weingartner:2001aa}. The remaining MCRT procedures follow standard practices, including sampling optical depths for scattering and iteratively updating the radiation field and ionization states until convergence. For detailed parameter choices, we follow a similar pipeline as described in \citet{Smith2022} and \citet{McClymont2024}.

After establishing the ionization equilibrium data, COLT also employs MCRT to model resonant lines from the FIRE-2 simulations by tracking individual photon packets and their interactions with ionic species.  For our analysis, to be consistent with the PCM and SALT assumptions, we consider only continuum scattering, assuming all photons originate from star particles adopting the same BPASS SEDs. Mock spectra are generated with the next-event estimation (peel-off) technique, collecting photon fluxes along six $\pm \hat{x}$, $\pm \hat{y}$, and $\pm \hat{z}$ sightlines recorded at two virial radii to ensure we capture the entire galaxy and its surrounding CGM.  The local velocity dispersion, which sets the spectral line profile, accounts for both microturbulent and thermal motions. While the latter can be obtained from the simulation, the former must be set by hand and we assume a uniform value of $10\rm\ km\ s^{-1}$ \citep{Chen2023}. 
Simulations were handled using YT \citep{Turk2011} under the FIRE-2 cosmology with dimensionless Hubble parameter $h = 0.702$, while data from \cite{Pandya2021} used $h = 0.6774$; when necessary we correct for this by rescaling all distances by the ratio $0.6774/0.702$.

\begin{figure*}
	\centering
	\includegraphics[scale=0.515]{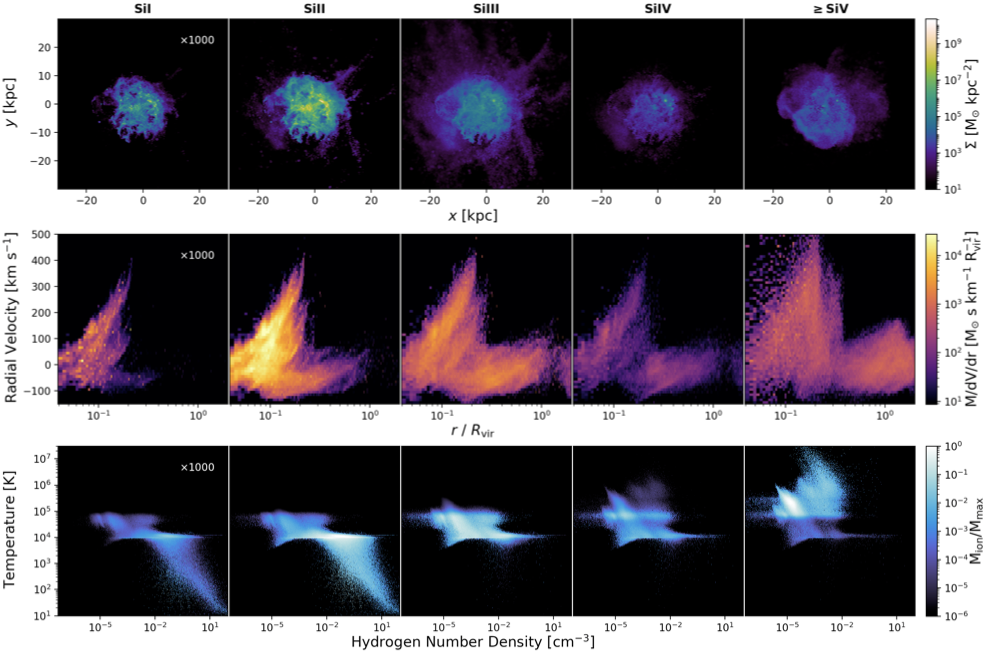}
	\caption{Various phase diagrams of the first five ionization states of silicon from simulation m11c at a local maximum in the mass outflow rate.  \emph{\textbf{Top Row}}: The surface density ($\Sigma$) is plotted according to spatial extent. $\rm Si^{+}$ and $\rm Si^{2+}$ account for the majority of matter, with $\rm Si^{0}$ shown in magnification.  \emph{\textbf{Middle Row}}: Particle mass per volume per radial projection is plotted according to radial velocity and radius normalized by the virial radius. Higher ionization states extend to larger radii and higher radial velocities than lower ionization states. \emph{\textbf{Bottom Row}} Ionic particle mass, normalized by the largest particle mass, is plotted against temperature and hydrogen number density. Only certain ionization states exist at the temperature extremes, but all ionization states are present between $10^4-10^5\ \rm K$. The hotter phases also appear as more diffuse gas.}
	\label{fig:phase_diagrams}
\end{figure*} 

In what follows, we demonstrate the COLT process by comparing our mock spectra to the CLASSY observations.  We focus on modeling the first five ionization states of silicon, where all higher ionization states are implicitly concentrated into $\rm Si^{4+}$.  On average, we find that $\rm Si^{+}$, $\rm Si^{2+}$, and $\rm Si^{3+}$ together already account for the majority ($>70\%$) of the total silicon mass under the COLT interpretation of FIRE-2. This finding supports previous works that use these ionization states to estimate the total silicon density (e.g., \citealt{Chisholm2018a,McKinney2019,Huberty2024}).  We conclude this section by designating the mock spectra to be used in our tests of outflow rate estimators.  


\subsection{The COLT Interpretation of FIRE-2}

In Figure~\ref{fig:phase_diagrams}, we present phase diagrams of the first five ionization states of silicon in the m11c simulation.  The chosen snapshot represents a local maximum in the mass outflow rate of an outflow episode at a lookback time of 0.25 Gyr.  

The top row shows the surface density according to spatial extent.  The different panels or phases demonstrate that the first two ionization states ($<\rm Si^{+}$) are more concentrated than the higher ionization states.  Note that the first ionization state has been magnified by a factor of 1000.  

The middle row shows the particle mass per velocity per virial radius as a function of radial velocity and radius.  In general, the galaxy has both inflowing ($v_r < 0$) and outflowing ($v_r >0$) gas with the outflows reaching much larger radial velocities than the inflows at all phases.  Furthermore, by comparing the different panels, it is clear that the higher ionization state gas ($> \rm Si^{+}$) can achieve much larger radial velocities than the lower ionization state gas.  These features are qualitatively in agreement with the spectral interpretation of the Si II 1190\AA, 1193\AA, 1260\AA, Si III 1206\AA, and Si IV 1394\AA, 1403\AA\ absorption lines in 10 Green Pea galaxies by \cite{Carr2021a}.  Note also the increasing prominence of inflows at larger radii presumably reflecting cosmological accretion or, in the case of highly ionized silicon, recycling of previous wind ejecta.  

The bottom row shows the particle mass of the relevant silicon ion according to temperature and number density of hydrogen.  The quantity has been normalized by the highest particle mass bin.  As expected, the higher ionization states occupy the largest temperature ranges.  A cutoff emerges around $10^5\rm\ K$ where the low ionization states ($\rm \leq Si^+$) do not exceed this value.  In addition, the higher ionization state particles ($\rm >Si^+$) appear to trace lower density material in terms of the hydrogen, and vice versa.  A similar scenario has been proposed for Mg II by \cite{Carr2024} who modeled the escape of ionizing radiation in multi-phase winds.  They proposed that Mg II traces cool dense clouds of hydrogen, which were embedded in a hotter lower density ambient medium.  These results are consistent with \cite{Decataldo2024} who studied the origin of cool gas in the CGM in the smoothed particle hydrodynamics cosmological zoom-in simulation Eris2k.  They attributed the abundance of cool gas at $\sim 10^4\rm\ K$ to photoionization equilibrium by the UV background and the build up of cool gas in the high density tail to shorter cooling times.

\begin{figure*}[htbp]
    \centering
    \subfigure{
        \includegraphics[width=0.485\textwidth]{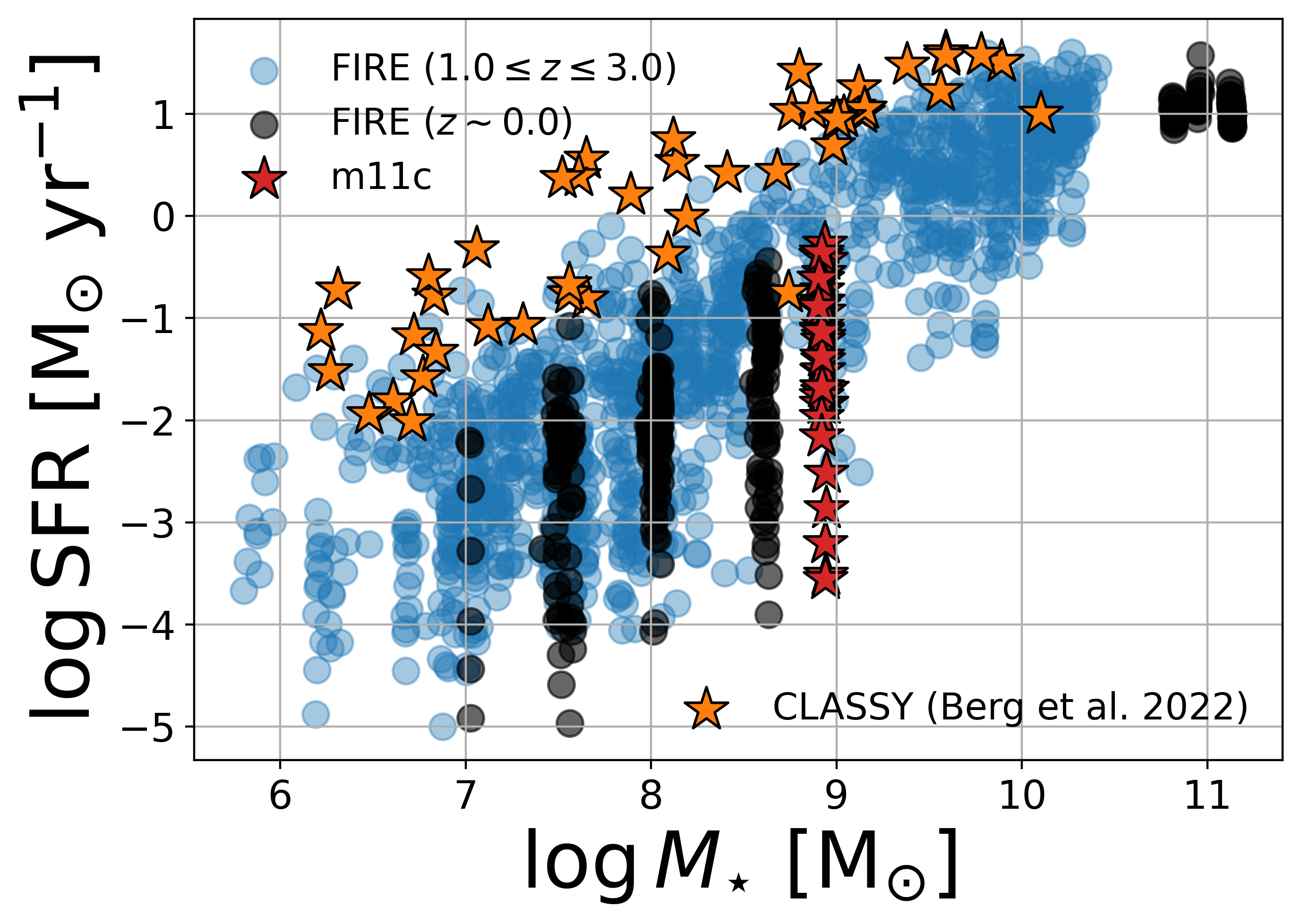}
    }
    \hfill
    \subfigure{
        \includegraphics[width=0.485\textwidth]{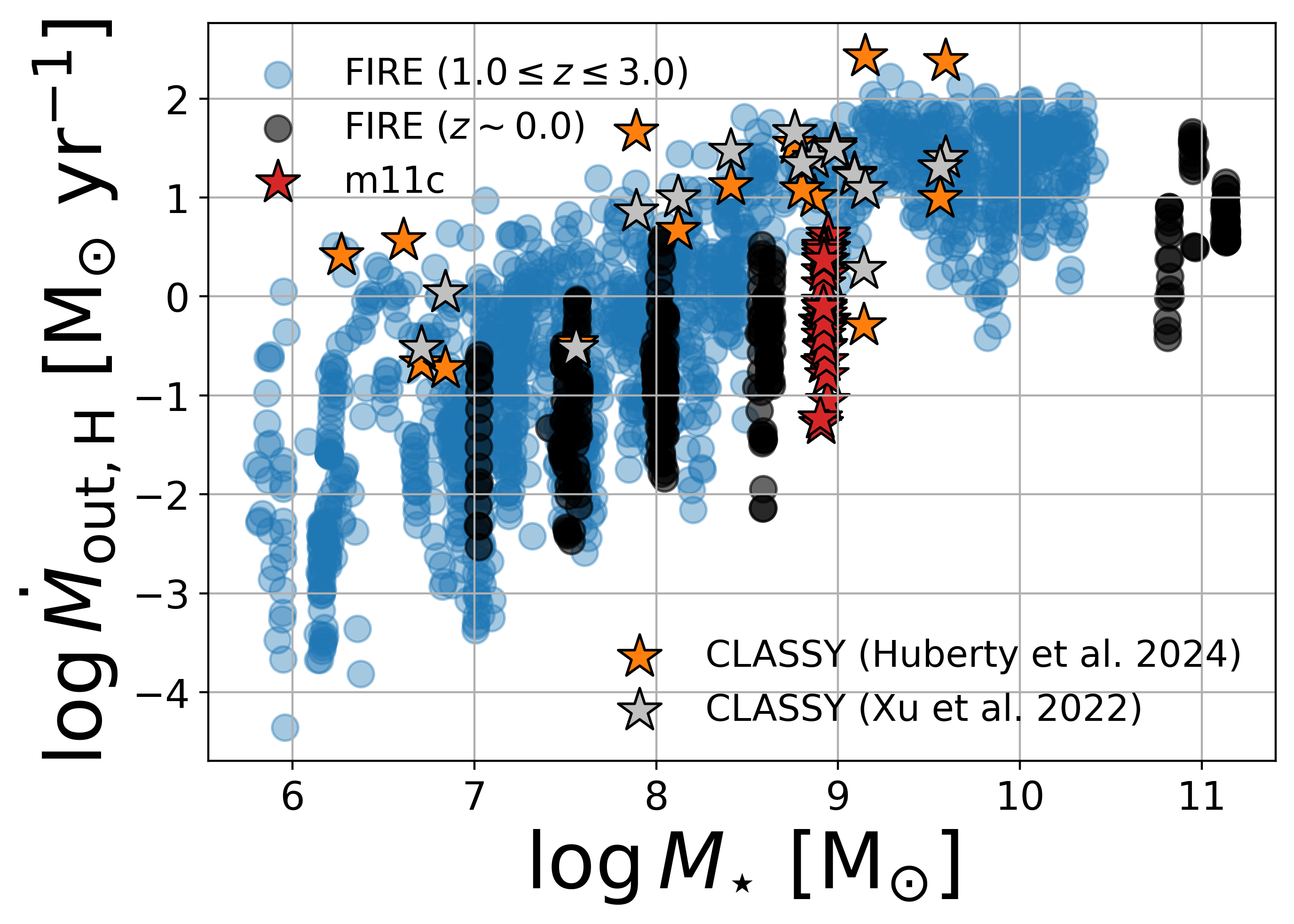}
    }
    \caption{A comparison of the properties of CLASSY galaxies to the corresponding FIRE-2 values measured at redshifts $z\sim 0$ (red) and $1\leq z \leq 3$ (light blue).  \emph{\textbf{Left Panel}}: The star formation rate (SFR) is shown as a function of stellar mass ($M_{\star}$). The CLASSY galaxies (gold stars) typically have higher SFRs than FIRE-2 galaxies (circles) at all redshifts, despite being local objects. \emph{\textbf{Right Panel}}: The mass outflow rate hydrogen ($\dot{M}_{\rm out, H}$) is shown as a function of $M{\star}$. The CLASSY values, measured by SALT (gold stars) and the PCM (silver stars), are comparable to the high redshift values in FIRE-2. The FIRE-2 values for simulation m11c are highlighted by red stars in each panel. We exclusively test the spectral predictions of SALT and the PCM with this simulation.}
    \label{fig:SFR}
\end{figure*}


\subsection{Comparison with CLASSY}
Since this study is motivated by the work of \cite{Huberty2024}, who highlighted differences in various model interpretations of the outflow properties in the CLASSY galaxies \citep{Berg2022}, we briefly compare the COLT interpretation of FIRE-2 to CLASSY.  Before comparing spectral line properties, we focus on how well the FIRE-2 simulations resemble the CLASSY galaxies.  We begin by comparing star formation rates (SFRs) and stellar masses, $M_{\star}$, in the left panel of Figure~\ref{fig:SFR}.  We consider snapshots at both local ($z\sim 0$, i.e., the values recorded in Table~\ref{tab:sims}) and at high redshifts ($1<z<3$).  We do this, because, although the CLASSY galaxies are local ($0.002 < z < 0.182$), they were selected as UV bright objects and were measured to have extremely high SFRs which better resemble objects at cosmic noon or around $z\sim 2$ \citep{Berg2022}.  Indeed, we find that the CLASSY SFRs agree better with the FIRE-2 values at higher redshifts, and even frequently exceed the highest values.  The CLASSY SFRs were measured from SED fitting using BEAGLE \citep{Chevallard2016} which models both stellar populations and nebular emission assuming a continuous SF history \citep{Berg2022}.  The FIRE-2 SFRs, on the other hand, are bursty (excluding the m12's), and our values represent averages measured over intervals of 20 Myrs.  While it is possible that using smaller time intervals could increase the values, it is unlikely to make up the difference \citep{Pandya2021}.  Thus, there is a real difference between the FIRE-2 simulations and CLASSY.  Given the novelty of the study, however, we simply note this caveat and continue our comparison.  We refer the reader to Figure 1 of \cite{Gandhi2022}, which compares the specific SFRs in FIRE-2 to a larger sample of observational surveys.             

In the right panel of Figure~\ref{fig:SFR}, we compare the hydrogen outflow rates, $\dot{M}_{\rm out, H}$, of CLASSY, estimated by \cite{Xu2022a} and \cite{Huberty2024}, to the FIRE-2 values as a function of $M_{\star}$.  The FIRE-2 values were measured in a shell of thickness $0.1-0.2R_{\rm vir}$ using Equation~\ref{eq:FIRE_MOR}.  Not surprisingly, the majority of estimates are in better agreement with the FIRE-2 values measured at high-redshift.  Unlike the SFR estimates, however, the majority of CLASSY MOR estimates do not eclipse the FIRE-2 values.  This may stem from the different radii used to estimate the MORs.  For example, \cite{Huberty2024} measured the MOR closer to the ISM at the UV half-light radius.  Another possibility is the incompatibility of SFR definitions.  For instance, \cite{Huberty2024} found a strong agreement between the mass loading factor versus $M_{\star}$ relationship in CLASSY with the extracted relationship from FIRE-2 by \cite{Pandya2021}.  \cite{Pandya2021} computed an instantaneous SFR by summing over the individual SFRs per particle in a shell of radius $0.1 R_{\rm vir}$.  Finally, it is possible that the CLASSY galaxies do indeed have lower mass outflow rates than FIRE, despite having higher SFRs.  This result is consistent with the preventative feedback scenario highlighted by \cite{Carr2023(Chris),Pandya2023,Voit2024a,Voit2024b}.  Overall, we find the qualitative agreement with $M_{\star}$ by both models and FIRE-2 to be encouraging.  Developing a better understanding of this relationship quantitatively is a main goal of this paper.  

\begin{figure*}[htbp]
    \centering
    \subfigure{
        \includegraphics[width=0.485\textwidth]{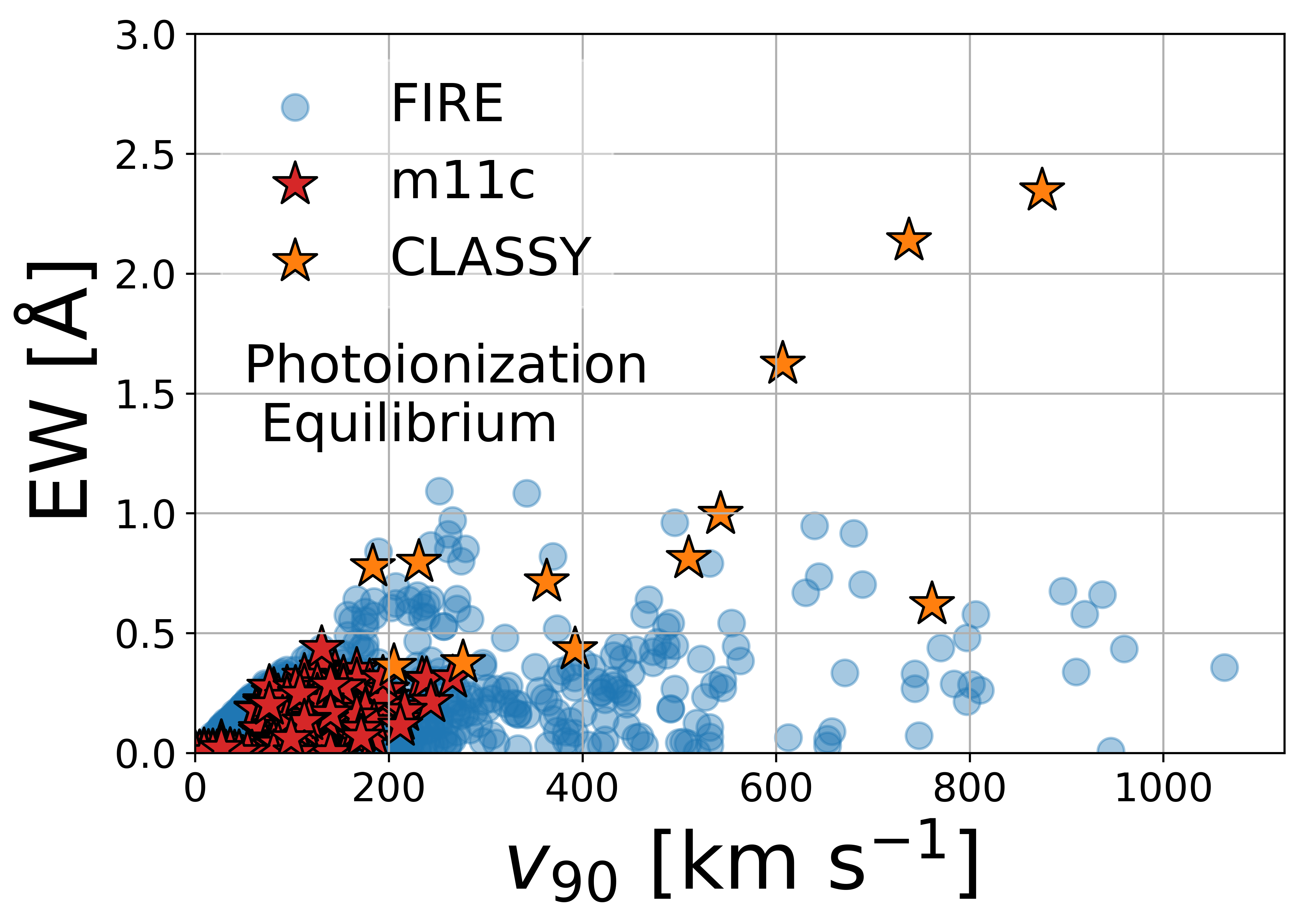}
    }
    \hfill
    \subfigure{
        \includegraphics[width=0.485\textwidth]{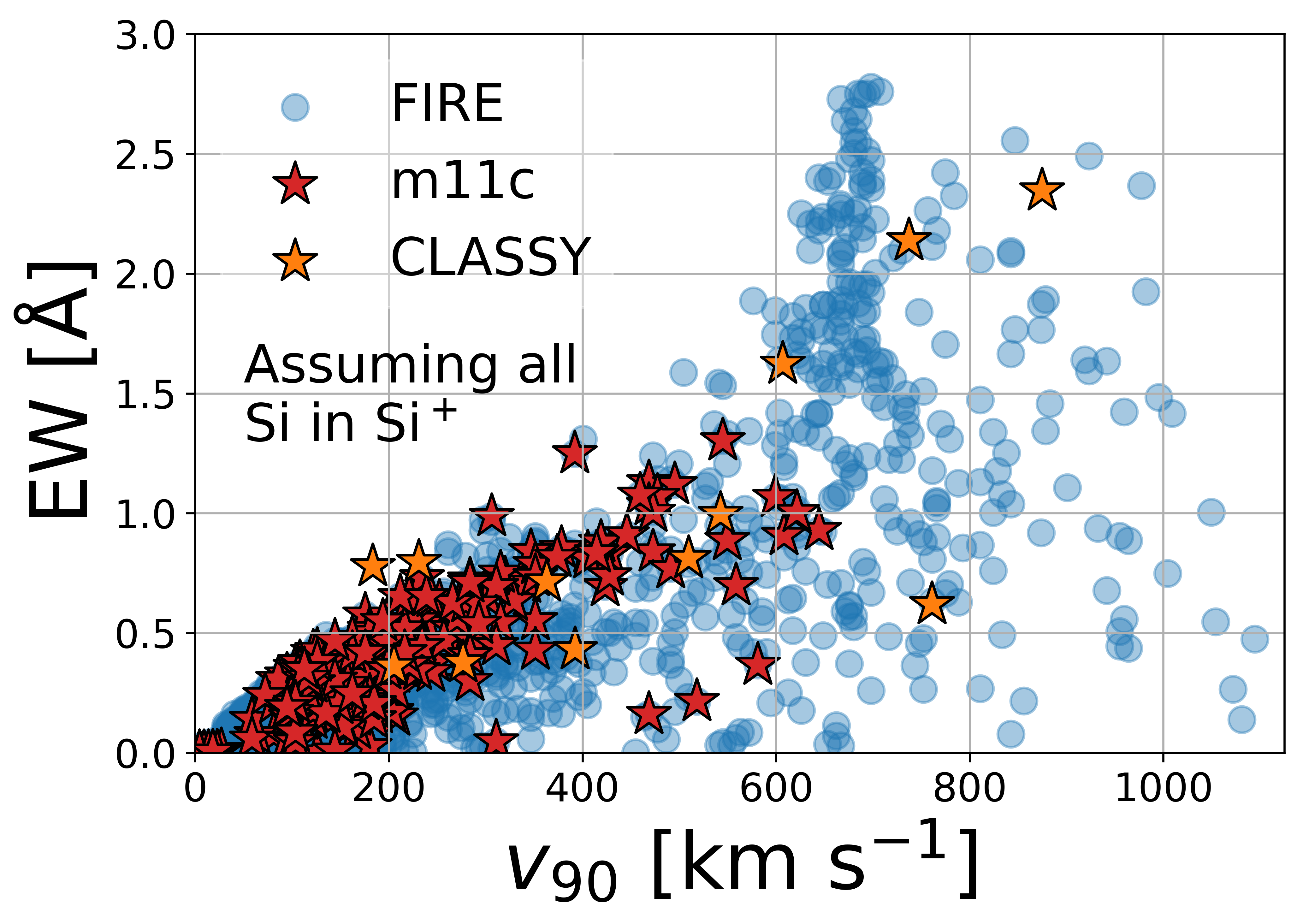}
    }
    \caption{A comparison of the equivalent widths (EWs) and $v_{90}$ values derived from the Si II 1190\AA\ lines of CLASSY (gold stars) to the corresponding FIRE-2 values (blue circles). The FIRE-2 values for simulation m11c are highlighted by red stars in each panel.  \emph{\textbf{Left Panel}}: Values taken from spectra computed using the full ionization treatment in COLT.  \emph{\textbf{Right Panel}}: Values derived from spectra assuming all silicon is in the first ionization state. This interpretation more accurately captures the features of the CLASSY data, which exhibit higher column densities and trace faster outflows.}
    \label{fig:EW_v90}
\end{figure*}

Having compared galaxy properties, we now move on to consider the properties of spectral lines.  For this aspect of the comparison, we generated spectra at local redshifts only, or from within the redshift ranges listed in Table~\ref{tab:sims}.  We assumed a resolution of 20 $\rm km\ s^{-1}$ and added Gaussian noise in post-processing to reach a signal-to-noise ratio of 10 in the continuum.  Given the vast differences in SFR and MOR between FIRE-2 and CLASSY at these redshifts, we suspect the wind features to be significantly different.  We investigate the severity by comparing equivalent widths (EWs) and $v_{90}$ values obtained from the Si II 1190\AA\ line in the left panel of Figure~\ref{fig:EW_v90}.  The $v_{90}$ values correspond to the observed velocity where the normalized area of the line profile is equal to 90\% of the EW when integrating left from line center (e.g., \citealt{Chisholm2017a}).  To ensure a fair comparison, we derived the EW and $v_{90}$ values directly from the SALT model fits to the spectra considering only the un-smoothed absorption component (see \citealt{Scarlata2015,Carr2018,Carr2023} for examples of pure absorption profiles generated with SALT).  In doing so, we are able to remove emission in-filling and the effects of instrumental smearing.  The CLASSY values were obtained from the model fits performed by \cite{Huberty2024} in 12/45 possible galaxies free of Milky Way contamination in the 1190\AA\ line.  The results show that the FIRE-2 outflows have much lower velocities and EWs than CLASSY.  We have checked, and the outflows with the largest $v_{90}$ values in FIRE-2 are found in the m12 simulations, which lie far outside the stellar mass range of CLASSY.  We have highlighted the values recovered from simulation m11c which represents the simulation with the largest stellar mass within the CLASSY sample.  The $v_{90}$ values for m11c rarely exceed 200 $\rm km\ s^{-1}$.  These low values may be the result of m11c's low metallicity, which is around $0.1Z_{\odot}$, and is barely within the range probed by CLASSY ($\sim 0.02-1.30Z_{\odot}$).  Hence, the winds of m11c likely undergo significantly less metal cooling than the CLASSY objects.  Lastly, we note that we generated spectra along only six LOS for each of the snapshots used in this comparison.  We ran a separate study with only m11c, considering 192 LOS, and found no major changes suggesting that we didn't “miss” the outflows by considering too few LOS.


Overall, the low EWs and $v_{90}$ values in the COLT interpretation of FIRE-2 at local redshifts prevent us from performing our comparison tests of the flow rate estimators. During trial runs, we found that outflows could not be identified in the vast majority of Si II absorption lines.  This occurs because the absorption features are not large enough to register detection of a net blue shift in absorption.  To remedy this situation, we decided to place all silicon into a single ionization state, namely $\rm Si^+$.  We show the new EWs and $v_{90}$ values in the right panel of Figure~\ref{fig:EW_v90}.  By forming winds composed of gas at all temperatures, we have increased the $v_{90}$ and EW values which are now in much better agreement with CLASSY.  The presence of these high velocity, hotter outflows supports our intuition that the CLASSY winds likely undergo substantially more metal line cooling than the winds of m11c, providing some physical justification for our decision to focus on a single ionization state.  We will use this interpretation of the data in the rest of our analysis.

Ultimately, this problem occurs because the CLASSY objects represent a population of exceptionally UV bright highly star forming objects and are essentially outliers.  The failure to find local representations of the CLASSY galaxies in simulation suites is not unique to FIRE-2 - for example, \cite{Gazagnes2023} and \cite{Jennings2025} opted to study the spectral features of CLASSY using RAMSES simulations at higher redshifts where the SFRs are inherently higher.  This alternative solution is not free of problems, however, since the physics of outflows and feedback (e.g., the mass-metallicity relation, \citealt{Erb2006}) can be expected to change with redshift.  Thus, both solutions are not without flaws.  We will discuss the limitations of our approach at the end of Section~\ref{sec:discussion}. 

\subsection{Mock Spectra of m11c}

To reduce our computational requirements, we limit our tests comparing the SALT and PCM spectra predictions to a single simulations, m11c.  We have highlighted the properties of this simulation in Figures~\ref{fig:SFR} and \ref{fig:EW_v90}.  We generate mock spectra along 48 different (HEALPix) lines of sight for 35 snapshots, spaced $\sim 24$ Myrs apart, creating a total of 1680 spectra.  We generate spectral lines for the Si II 1190\AA, 1193\AA, 1260\AA, 1304\AA, 1527\AA, or the same set of lines studied by \cite{Huberty2024}.  These lines were not special to this study, however, and are frequently used to constrain the cool phase of the CGM (e.g., \citealt{Henry2015,Carr2021a,Berg2022}).  In addition, with an ionization potential of  of 16.3 eV, $\rm Si^{+}$ has been used as a low-ionization state (LIS) tracer to map the neutral gas content of the CGM to predict the LyC escape fraction \citep{Chisholm2017a}.    

These lines cover a range of oscillator strengths, with the most optically thin line being 1304\AA.  As the PCM is known to perform better under optically thin conditions \citep{Savage1991}, this should give it the best opportunity to succeed given plausible observing conditions.  The atomic structure of $\text{Si}^{+}$ admits both resonant and fluorescent transitions for each of the resonant lines we are considering (see \citealt{Scarlata2015} for a diagram illustrating the 1190\AA, 1193\AA\ doublet and its associated fluorescent transitions).  COLT can accommodate both types of emission, but among the interpretive models, only SALT accounts for the emission component; the PCM does not.  While we consider the specific atomic conditions governing the radiation transfer of the lines, we acknowledge that we do not attempt to simulate the full range of observational conditions. For example, we exclude the effects of blending with other metal lines and masking by Milky Way absorption (e.g., \citealt{Xu2022a}).  All relevant atomic information has been provided in Table~\ref{tab:atomicdata}.

\begin{table*}[t]
\caption{Atomic data for various lines of Si II.  Data taken from the NIST Atomic Spectra Database$^{\MakeLowercase{a}}$.}
\resizebox{2.1\columnwidth}{!}{$\begin{tabular}{c|c|c|c|c|c|c|c}\hline\hline
Ion & Vac. Wavelength&  $A_{ul}$&$f_{ul}$& $E_{l}-E_{u}$&$g_l-g_u$&Lower Level&Upper Level\\
&\AA& $s^{-1}$ &&$eV$&&Conf.,Term,J&Conf.,Term,J\\[.5 ex]
\hline 
 Si II& $1190.416$ &$6.53\times 10^{8}$&$0.277$&$0.0000000-10.41520 $&$2-4$ &$3s^23p, {}^2P^0, 1/2$&$3s3p^2, {}^2P, 3/2$ \\
Si II&$1193.290$ & $2.69\times 10^{9}$ &$0.575$&$0.0000000-10.390118$&$2-2$&$3s^23p, {}^2P^0, 1/2$&$3s3p^2, {}^2P, 1/2$  \\
$\rm Si\ II^*$& $1194.500$ & $3.45\times 10^{9}$ &$0.737$&$0.035613-10.41520$&$4-4$&$3s^23p, {}^2P^0, 3/2$&$3s3p^2, {}^2P, 3/2$ \\
$\rm Si\ II^*$& $1197.394$ & $1.4\times 10^{9}$ &$0.15$&$0.035613-10.390118$&$4-2$&$3s^23p, {}^2P^0, 3/2$&$3s3p^2, {}^2P, 1/2$ \\
\hline 
Si II&$1260.422$ & $2.57\times 10^{9}$ &$1.22$&$0.0000000-9.836720$&$2-4$&$3s^23p, {}^2P^0, 1/2$&$3s^23d, {}^2D, 3/2$  \\
$\rm Si\ II^*$&$1265.002$ & $4.73\times 10^{8}$ &$0.113$&$0.035613-9.836720$&$4-4$&$3s^23p, {}^2P^0, 3/2$&$3s^23d, {}^2D, 3/2$  \\
\hline 
Si II&$1304.370$ & $3.64\times 10^{8}$ &$0.0928$&$0.000000-9.505292$&$2-2$&$3s^23p, {}^2P^0, 1/2$&$3s3p^2, {}^2S, 1/2$  \\
$\rm Si\ II^*$&$1309.276$ & $6.23\times 10^{8}$ &$0.08$&$0.035613-9.505292$&$4-2$&$3s^23p, {}^2P^0, 3/2$&$3s3p^2, {}^2S, 1/2$  \\
\hline 
Si II&$1526.707$& $3.81\times 10^{8}$ &$0.133$&$0.0000000-8.121023$&$2-2$&$3s^23p, {}^2P^0, 1/2$&$3s^24s, {}^2S, 1/2$  \\
$\rm Si\ II^*$&$1533.431$ & $7.52\times 10^{8}$ &$0.133$&$0.035613-8.121023$&$4-2$&$3s^23p, {}^2P^0, 3/2$&$3s^24s, {}^2S, 1/2$  \\
[1ex]
\hline
\end{tabular} $}
\label{tab:atomicdata}
\\$^{\text{a}}$ http://www.nist.gov/pml/data/asd.cfm
\end{table*}

In Figure~\ref{fig:geometry}, we show stereographic projections or all sky maps of the mass-weighted radial velocities of silicon for each snapshot.  The time evolution of the sky maps is plotted according to $\dot{M}_{\rm{out},Z}^F$ measured at $0.15R_{\rm vir}$ as a function of lookback time.  Note there is a lag between the starburst and peak of the outflow rate \citep{Pandya2021}.  Each image is oriented so that the net angular momentum of the stars points from north to south. Outflowing gas is shown in red and inflowing gas in blue.  The bottom row provides possibly the easiest sequence to visualize: An outburst occurs which grows into a large spherical outflow by the time the outflow rate reaches a maximum. This is evidenced by the solid red sky maps. The outflow rate then begins to wane and recycling occurs.  During this time, inflow rates start to dominate as evidenced by the emergence of the blue regions. Eventually another outburst occurs. In this instance, the ensuing flow appears to be bi-conical as evidenced by the two red dots beginning to appear. In summary, the snapshots span roughly one Gyr in lookback time and cover roughly three outflow episodes or baryonic cycles, which include both inflows and outflows.  This time span will allow us to test both SALT and the PCM under some of the most realistic galaxy conditions currently obtainable by high-resolution cosmological hydrodynamics simulations of galaxy evolution.

\subsection{Identifying Outflows}

\begin{sidewaysfigure*}
    \centering
    \vspace{-10cm}
    \includegraphics[width=\textwidth,keepaspectratio]{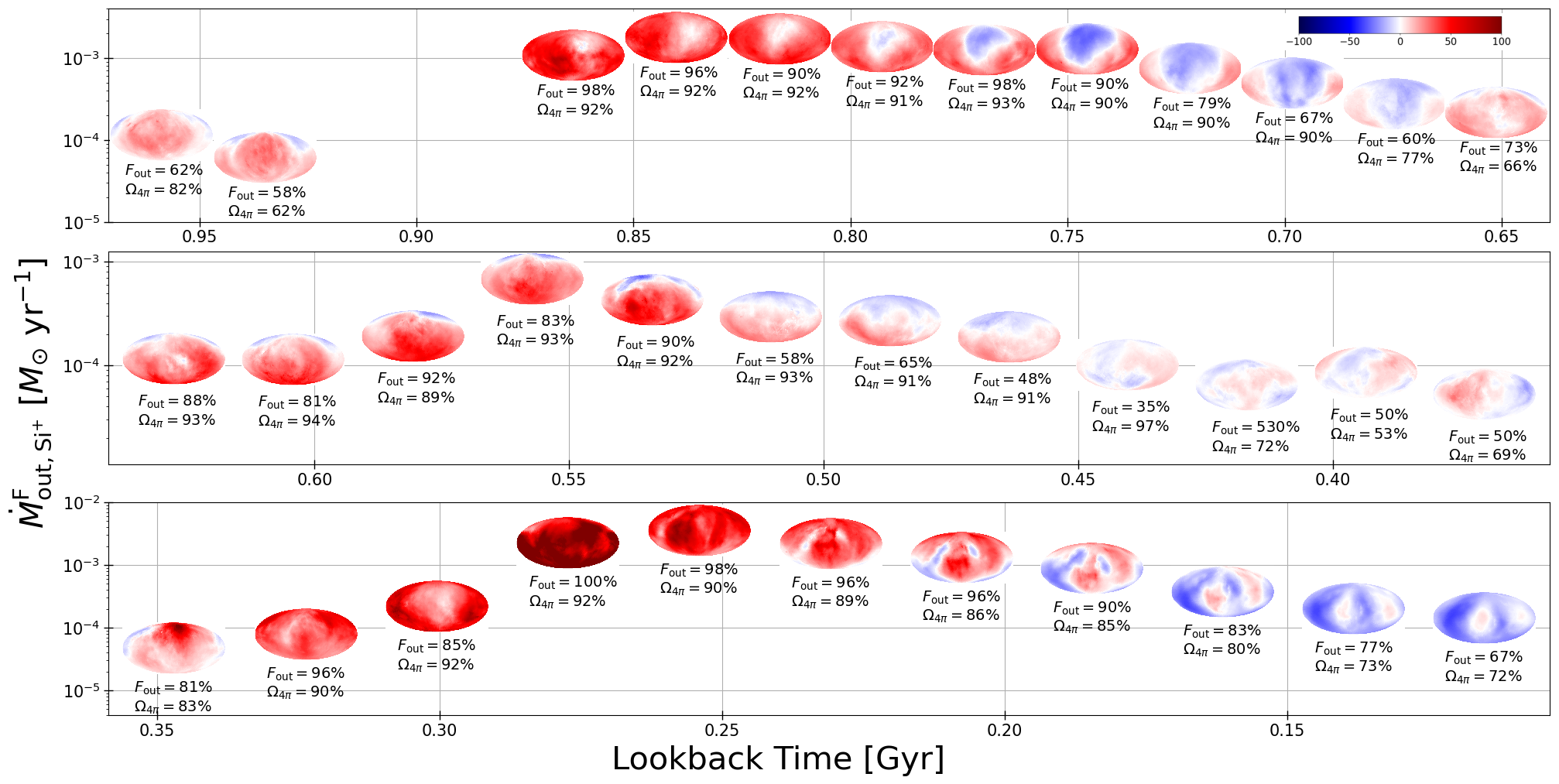}
    \caption{Stereographic projections or all-sky maps of the net radial outflow velocity as a function of lookback time. Each map is plotted according to the outflow rate measured at 0.15 $R_{\rm vir}$. Blue regions indicate inflows, while red regions indicate outflows. Below each sky map, the fraction of lines of sight (LOS) showing signs of outflowing gas in absorption according to the F-test, $F_{\rm out}$, are listed.  $F_{\rm out}$ tends to increase and decrease with the outflow rate matching the color distribution of each sky map.  In addition, we list the median solid angle, deduced from the different lines of sight by SALT, normalized by $4\pi$, $\Omega_{4\pi}$, beneath each map.  $\Omega_{4\pi}$ tends to trace $F_{\rm out}$, indicating that SALT is able to recover meaningful information about the gas geometry from the spectra.  }
    \label{fig:geometry}
\end{sidewaysfigure*}

To adhere to standard observational procedures, we perform an F-test before fitting to establish the presence of outflows.  We use the same procedure as \cite{Xu2022a}, discussed in section~\ref{sec:Models}, applied to the Si II 1260\AA\ line, or the line profile with the largest oscillator strength-wavelength product.  We compute the F-value, $F$, as 
\begin{eqnarray}
    F = \frac{\chi_{\rm ISM}^2-\chi_{\rm outflow}^2}{\chi_{\rm outflow}^2}\frac{n-p_{\rm outflow}}{p_{\rm outflow}-p_{\rm ISM}},
    \label{eq:f-value}
\end{eqnarray}
where $\chi_{i}^2$ is the chi-squared value for the appropriate model, $p_{\rm outflow} = 7$ and $p_{\rm ISM} = 2$ are the number of free parameters for the PCM and ISM models, respectively, and $n$ is the number of observed velocity bins used to calculate the line profile.  We reject the null hypothesis, i.e., that an outflow was not detected, at a desired false-rejection probability $\alpha = 0.05$, in agreement with \cite{Xu2022a}.  All spectra which fail the F-test were removed from further analysis.   We list the fraction of sightlines which detected an outflow, $F_{\rm out}$, beneath each all sky map in Figure~\ref{fig:geometry}.  In general, the pictures align with our intuition for $F_{\rm out}$: as the surface area of the outflow increases, so should $F_{\rm out}$.  The fact that this occurs in tandem with $\dot{M}_{\rm{out},Z}^F$ suggests that the solid angle of the outflow is initially small but then widens as the starburst continues to evolve.

\begin{figure*}
	\centering
	\includegraphics[scale=0.475]{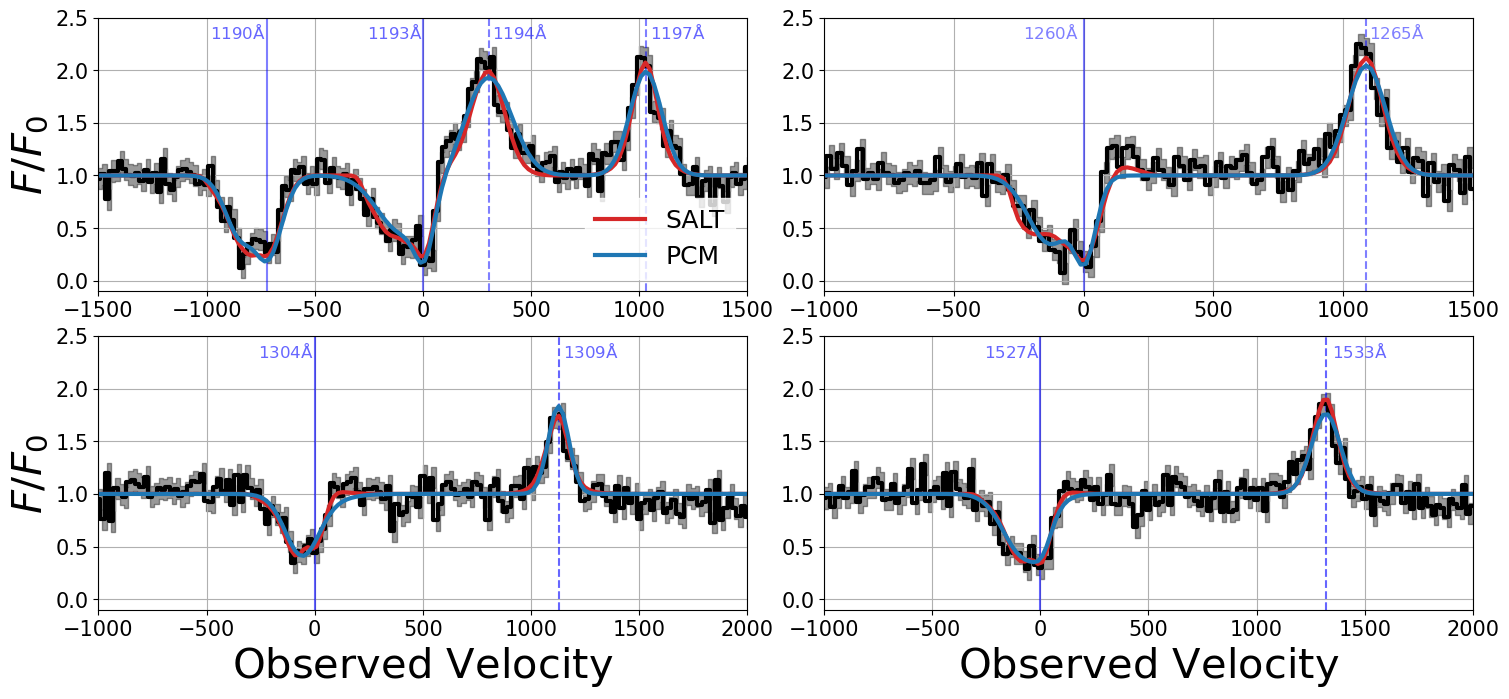}
	\caption{Mock Si II 1190\AA, 1193\AA, 1260\AA, 1304\AA, and 1527\AA\ spectral lines (black) generated from the m11c simulation using COLT, along with the best-fitting SALT model (red) and PCM (blue).  The chosen snapshot corresponds to a local maximum in $\dot{M}^F_{\rm out,Si^+}$ from an arbitrarily chosen LOS.  Each spectrum was smoothed with a Gaussian kernel to a resolution of $20\rm\ km\ s^{-1}$, and Gaussian noise was added until the continuum achieved a $S/N = 10$.  In other words, the fluctuations are either real features or added noise, as opposed to Monte Carlo noise.  The spectra exhibit both prominent emission and absorption features, with the blue-shifted asymmetric absorption trough indicating the presence of outflowing gas.}
	\label{fig:best_fit}
\end{figure*}
   
\subsection{Model Fitting}

We simultaneously fit all the aforementioned $\text{Si}^+$ lines using both the PCM and SALT models, using the same Bayesian-based fitting procedure as described in \cite{Carr2021a}.  In summary, we sample the natural log of a Gaussian likelihood function using the Python package, \texttt{emcee} \citep{Foreman-Mackey2013}, which relies on the Python implementation of Goodman's and Weare's Affine Invariant Markov Chain Monte Carlo (MCMC) Ensemble sampler \citep{Goodman2010}.  To ensure convergence of the posterior distributions, we explored the parameter space of each model with twice as many walkers as free parameters (i.e., $2\times[10+2 \times\#\ \rm{resonant\ lines}] = 40$ for SALT and $2\times [5\times \#\ \rm{resonant\ lines}] = 50$ for the pre-Gaussian fits plus $4$ from the minimization procedure for the PCM) for 10000 steps.  We removed the first 4000 steps, or the so-called “burn-in period” from each chain to ensure convergence.  We sample uniform priors and draw from the same initial conditions defined in \cite{Huberty2024}.  Each model was smoothed to a resolution of $20\rm \ km\ s^{-1}$ using SciPy's Gaussian filter before performing the model fits.


We employ two statistics to determine the “best-fit”: the maximum likelihood estimate (MLE) and the median of the posterior distributions (noting that the mean and mode yield similar results). For dependent quantities, such as mass outflow rates, we create histograms from the relevant SALT parameter chains and apply the same statistics to find the best fit. Figure~\ref{fig:best_fit} shows the fits based on the MLE for both SALT and PCM models.


\section{Results}
\label{sec:results}

The mass, momentum, and energy outflow rates are fundamental for assessing feedback efficiency and its influence on galaxy evolution. This is particularly evident in the FIRE-2 simulations, as shown by \cite{Pandya2021} and \cite{Pandya2023}.  Column densities and other properties such as the velocity structure of lines also offer insight into the inner workings of galaxy formation and are essential for inferring quantities such as escape fractions.  Absorption line studies offer an important and widely used means for extracting these properties from outflows.  However, as highlighted by \cite{Huberty2024}, the various models and methods used for interpreting spectra lines can report vastly different values.  This problem manifests from the different assumptions and limitations characteristic to each model, as well as a disagreement between base definitions.  For instance, without constraints on the underlying density and velocity field, most models typically assume a constant flow rate, whose value will depend on unconstrained quantities such as the fundamental scale of the outflow.  The primary goal for this paper is to test outflow rate estimators against simulations to better understand the meaning of their assumed definitions and uncertainties.  In this section, we test the ability of SALT and PCM to predict the outflow properties of the FIRE-2 simulations from mock spectra using standard observational procedures.  In doing so, we hope to point the field in a common direction in order to establish a self-consistent framework for estimating outflow rates, with known biases and uncertainties, which will allow for an unambiguous comparison to simulations and theoretical investigations.

\begin{figure*}
	\centering
	\includegraphics[width=\textwidth]{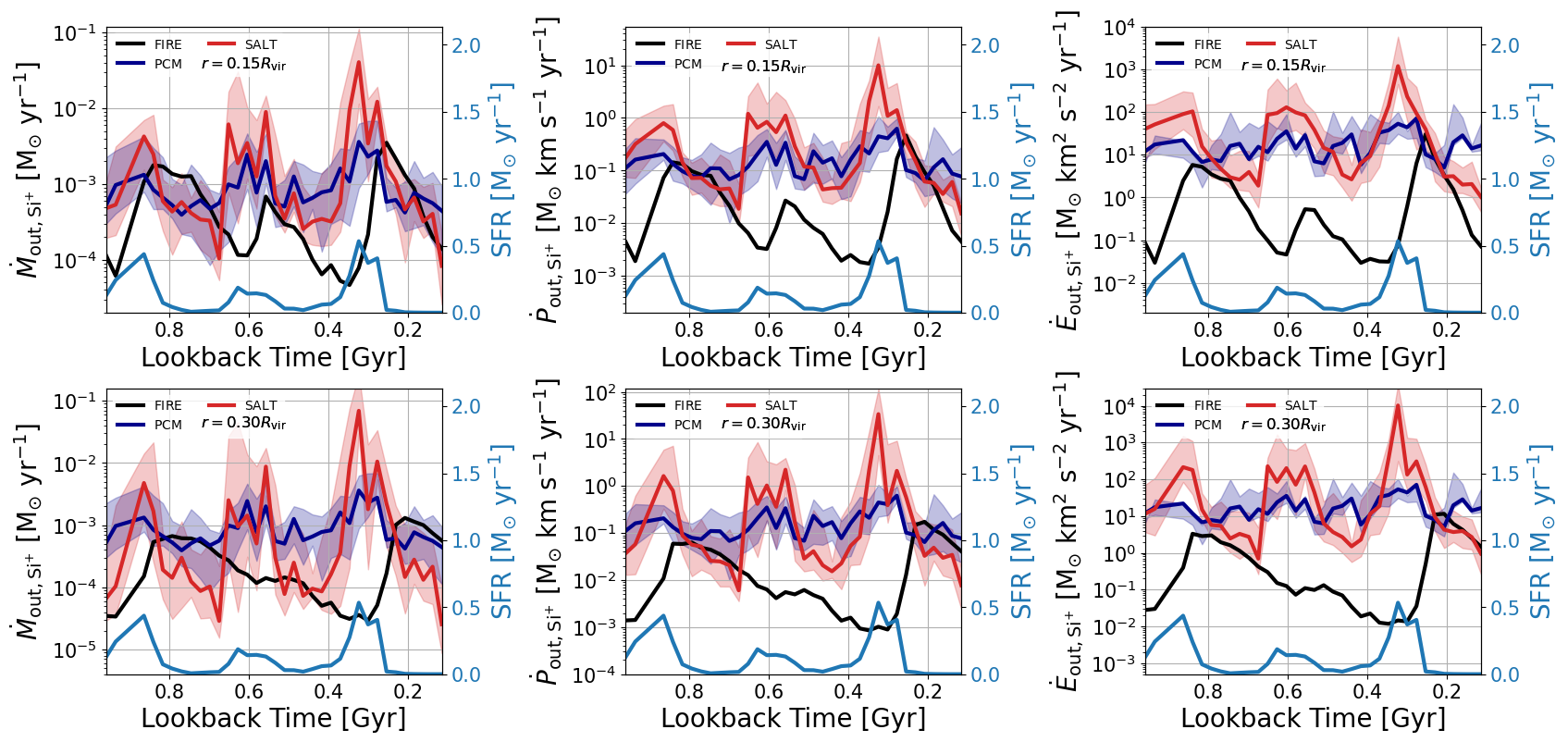}
	\caption{From left to right: SALT (red) and PCM (blue) estimates of $\dot{M}{\rm out,Si^+}$, $\dot{P}{\rm out,Si^+}$, and $\dot{E}{\rm out,Si^+}$ compared to FIRE-2 values (black) as functions of lookback time. Values are measured at $r = 0.15 \ R{\rm vir}$ (\textbf{\emph{Top Row}}) and $r = 0.30 \ R_{\rm vir}$ (\textbf{\emph{Bottom Row}}). Estimates represent the median across all LOS, with error bars indicating the median deviation above and below these values. The star formation rate (SFR, light blue) is shown at the bottom of each panel, with scale on the right vertical axis. PCM predictions show little time variation and typically overestimate FIRE-2 values. SALT agrees best with FIRE-2 near local maxima in flow rates and local minima in SFR but overestimates FIRE-2 otherwise. The origins of these trends are explored in subsequent figures and discussion.}
	\label{fig:mor_vs_time}
\end{figure*} 

    \subsection{Time Dependence of Flow Rates}

We begin our analysis with the time evolution of $\dot{M}_{\rm out, Si^+}$, $\dot{P}_{\rm out, Si^+}$, and $\dot{E}_{\rm out, Si^+}$ shown in the left, middle, and right columns of Figure~\ref{fig:mor_vs_time}, respectively.  We also include the SFR, averaged over 20 Myr, for reference.  The values in the top row were measured at $r = 0.15R_{\rm vir}$ and the bottom row at $r = 0.3R_{\rm vir}$.  The FIRE-2 values were obtained from a shell of thickness $0.1R_{\rm vir}$, centered at each radius.  The recorded values represent the median value, inferred from the LOS distribution for each snapshot.  There are at most 48 values with potential losses to LOS that failed the F-test.  The errors correspond to the median of the deviation from above and below these values, consistent with \cite{Carr2021a}.  As a reminder, the PCM assumes a constant flow rate.  

The PCM shows a slight evolution in the three measured quantities, though there may be some dependence on SFR, with a variation in measurements spanning 0.55 dex on average for each quantity. While we will explore the underlying cause of this behavior later, it is worth noting that the spectra show significant temporal variation. In general, PCM tends to overestimate the flow rates but aligns more closely with $\dot{M}^F_{\rm out, Si^+}$ than with $\dot{P}^F_{\rm out, Si^+}$ and $\dot{E}^F_{\rm out, Si^+}$.

SALT estimates, on the other hand, display periodic variations but do not consistently align with FIRE-2 values across all radii. The best agreement between SALT and FIRE-2 occurs in the ranges 0.8–0.65 Gyr, 0.55–0.45 Gyr, and 0.25–0 Gyr for $\dot{M}^F_{\rm out, Si^+}$ and in the ranges 0.8–0.65 Gyr and 0.25–0 Gyr for $\dot{P}^F_{\rm out, Si^+}$ and $\dot{E}^F_{\rm out, Si^+}$. Understanding these trends is a key objective of this study, but we note here that these periods correspond to local maxima in the FIRE-2 flow rates, with the smallest value occurring within the middle range. Moreover, SALT performs its worst when the SFR approaches local maxima.  

In ranges 0.8-0.65 Gyr, 0.55-0.45 Gyr and 0.55-2.50 Gyr, we measure the mean deviations from the FIRE-2 values for $\dot{M}_{\rm out, Si^+}$, $\dot{P}_{\rm out, Si^+}$, $\dot{E}_{\rm out, Si^+}$ of 0.36 (0.63), 0.56 (0.56) and 0.97 (0.80) dex at $0.15(0.30)R_{\rm vir}$, respectively.

\subsection{Radial Dependence of Flow Rates}

\begin{figure*}
	\centering
	\includegraphics[width=\textwidth]{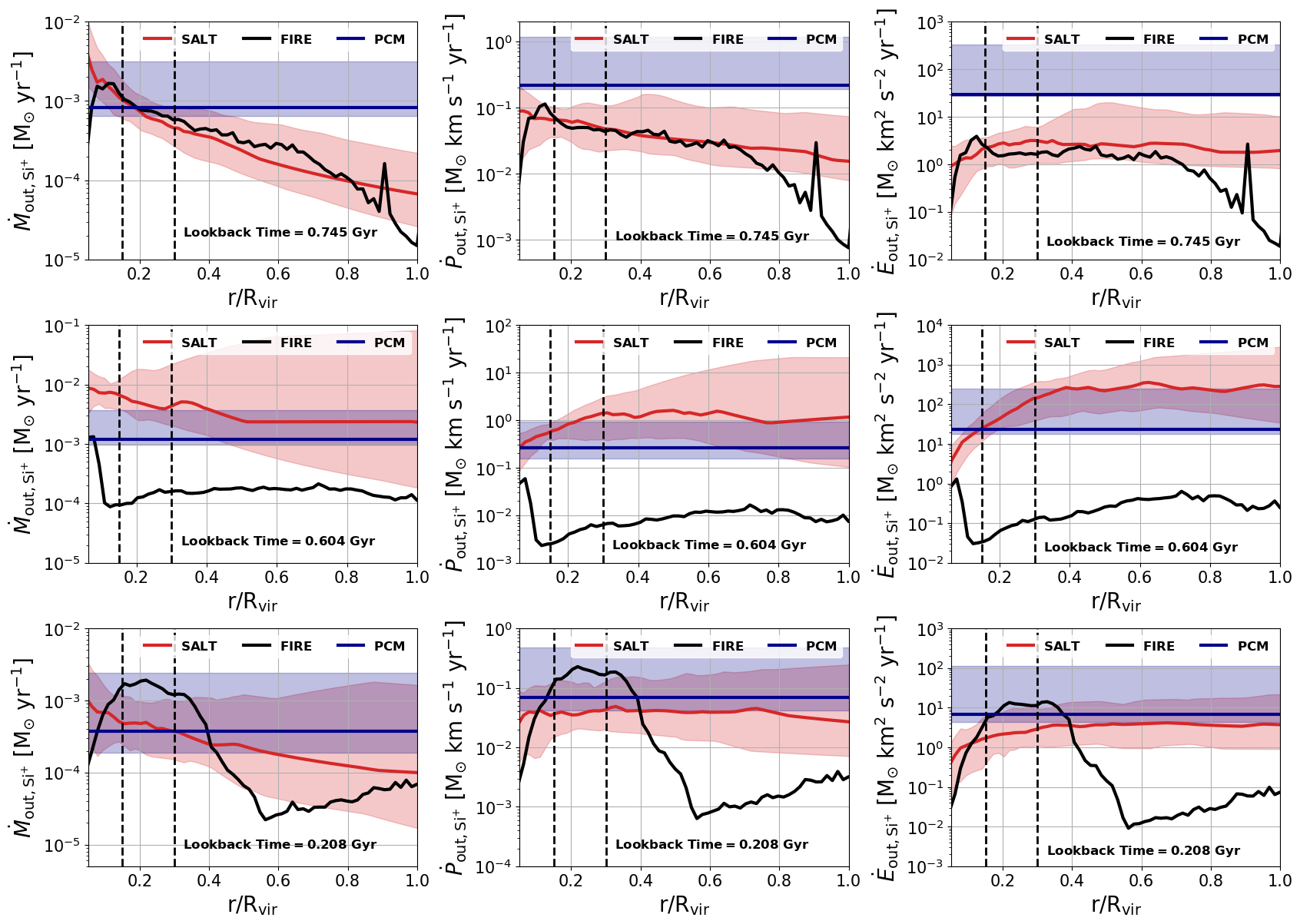}
	\caption{Shown from left to right: The SALT estimates (red) of $\dot{M}_{\text{out,Si}^+}$, $\dot{P}_{\text{out,Si}^+}$, and $\dot{E}_{\text{out,Si}^+}$ as functions of radius, compared to the FIRE-2 values (black), for fixed snapshots at lookback times of 0.745, 0.604, and 0.208 Gyr in the \textbf{\emph{Top}}, \textbf{\emph{Middle}}, and \textbf{\emph{Bottom Rows}}, respectively. Estimates represent the median over all LOS, and the error bars represent the median of the deviation from above and below these values.  The radially constant value predicted by the PCM is shown in dark blue.  The horizontal axis begins at $r=R_{\rm SF}$ and the dashed lines correspond to $0.15$ and $0.3R_{\rm vir}$, or the distances corresponding to the measurements used in Figure~\ref{fig:mor_vs_time}.  In general, the flow rates in FIRE-2 are not well described by simple power laws.  The top row represents an instance where SALT succeeds, the middle row where it cannot describe the behavior of the flow, and the bottom row where it can only capture the behavior of the flow rates over a portion of $R_{\rm vir}$ -- although, the median quantity appears to fail to capture the behavior of $\dot{M}^F_{\rm out,Si^+}$ and $\dot{P}^F_{\rm out,Si^+}$ over the full range.  }
	\label{fig:mpe_radial}
\end{figure*}

We show the radial dependence of $\dot{M}_{\rm Si^+}$, $\dot{P}_{\rm Si^+}$, and $\dot{E}_{\rm Si^+}$ for different snapshots in the left, middle, and right columns of Figure~\ref{fig:mpe_radial}, respectively. The top, middle, and bottom rows correspond to lookback times of 0.745, 0.604, and 0.208 Gyr, respectively. To derive the radial profiles for SALT, we computed the MLE at each radius and then took the median value from the 48 different lines of sight.  Once again, we discard LOS that failed the f-test.  The errors represent the median of the absolute deviation from above and below the median values. We also show the PCM predictions for reference, as it does not account for radial dependence.  As we will see, the actual radial locations where the PCM agrees with FIRE-2 are likely fortuitous. 

In general, the FIRE-2 flow rates are non-constant and show a variety of shapes which can deviate from simple power-laws.  We have selected snapshots to highlight instances where SALT succeeds and fails to capture this behavior. The top row displays flow rates which are well captured by SALT and follow a power-law structure.  In contrast, the middle row shows a rapidly decaying flow rate which quickly becomes constant.  SALT appears to dramatically overestimate the flow rates when this is the case.  The bottom row shows a scenario where the flow rates initially increase, flatten out, and then decay.  SALT can only conceivably capture a given portion of this behavior.  However, SALT does not appear to describe the FIRE-2 values over any specific range, although it performs best for $\dot{E}_{\rm out,Si^+}^F$ at small radii.  We will explore how this behavior depends on the SALT predictions of the density and velocity fields in the next sub-section. 

The variation of the PCM estimates carry over from Figure~\ref{fig:mor_vs_time}.  The respective variations on $\dot{M}^S_{\rm Si^+}$, $\dot{P}^S_{\rm Si^+}$, and $\dot{E}^S_{\rm Si^+}$, averaged over radius, are 0.51, 0.70, 0.67 dex, 1.5, 1.1, 0.91 dex, and 1.1, 1.0, 0.78 dex for lookback times of 0.745, 0.604, and 0.208 Gyr, respectively.

\begin{figure*}
	\centering
	\includegraphics[width=\textwidth]{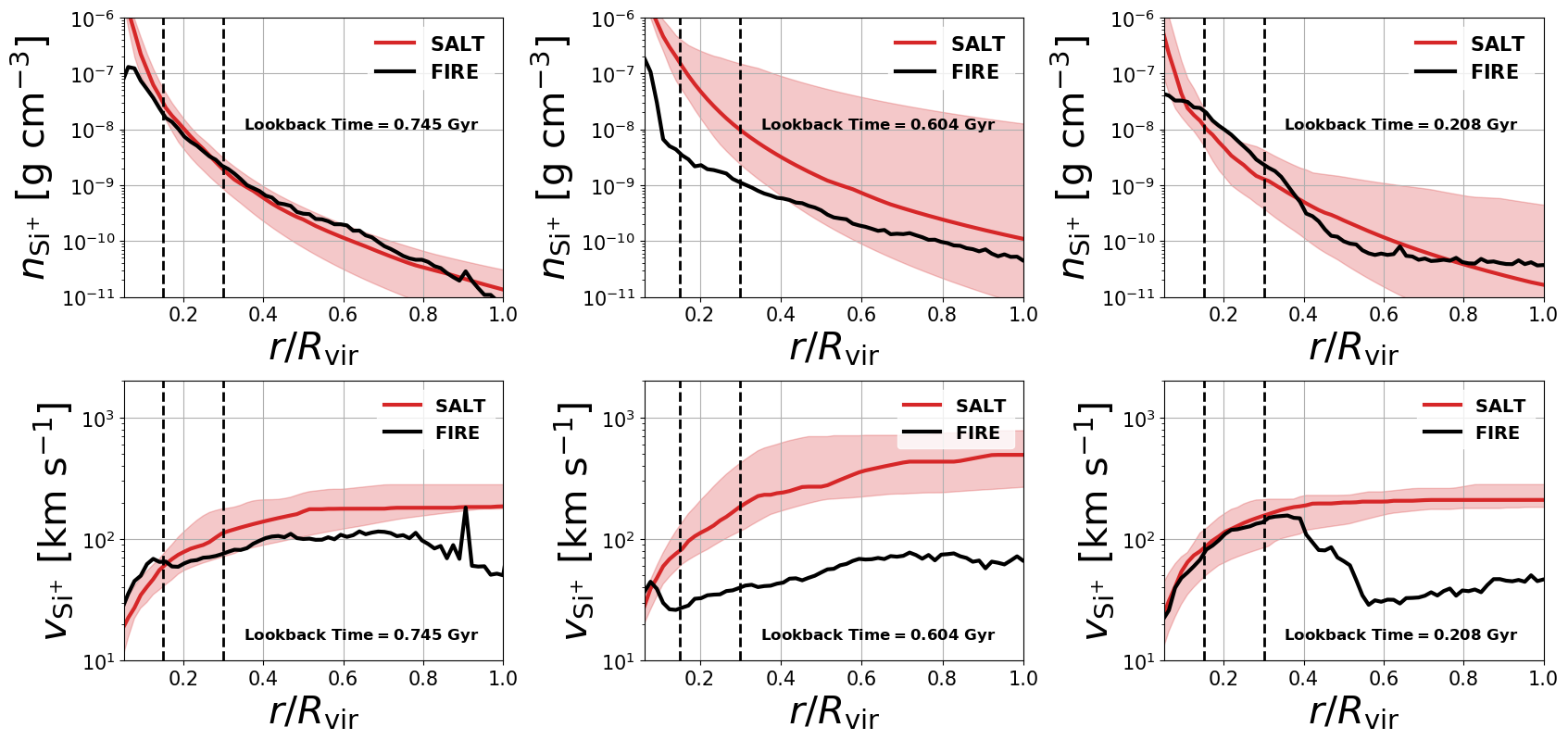}
	\caption{Comparison of SALT predictions (red) to FIRE-2 values (black) for the density (\textbf{\emph{Top Panels}}) and velocity fields (\textbf{\emph{Bottom Panels}}). The \textbf{\emph{Left}}, \textbf{\emph{Middle}}, and \textbf{\emph{Right Columns}} correspond to lookback times of 0.745, 0.604, and 0.208 Gyr, aligning with the rows in Figure~\ref{fig:mpe_radial}. Estimates represent the median across all LOS, with error bars indicating the median deviation above and below these values. SALT accurately predicts both fields at 0.745 Gyr but fails to capture their behavior at 0.604 Gyr. At 0.208 Gyr, SALT reproduces the velocity field at $r < 0.4 R_{\rm vir}$ but not the density field. These discrepancies influence the flow rate predictions shown in Figure~\ref{fig:mpe_radial}.}
	\label{fig:density_velocity}
\end{figure*}

\subsection{Density and Velocity Fields}

We further investigate the radial dependence of the flow rates by comparing the SALT estimates of the density and velocity fields to the FIRE-2 values in the top and bottom rows of Figure~\ref{fig:density_velocity}, respectively.  The left, middle, and right columns correspond to snapshots at lookback times of 0.745, 0.604, and 0.208 Gyr, respectively, and correspond to the rows in Figure~\ref{fig:mpe_radial}.  Once again, the radial profiles were computed by taking the median of the MLE at each radius and the errors represent the median of the absolute deviation from above and below this value.  LOS which failed the F-test were removed.

Of the three snapshots, SALT performed best at 0.745 Gyr, accurately capturing both the velocity and density fields. Consequently, it also reproduced the flow rates at this lookback time in Figure~\ref{fig:mpe_radial}. However, at 0.604 Gyr, SALT failed to capture both fields. In this case, the velocity field is notably shallow and falls outside the prior for $\gamma$, the power-law index of the velocity profile (see Table~\ref{tab:salt_free_parameters}). As a result, SALT dramatically overestimates the velocity field at nearly all radii, leading to a similarly large overestimate of the density field. The combined effect inflates the predicted flow rates by orders of magnitude.

Spectrally, this behavior would correspond to a deep and broad absorption profile for a spherical outflow (see \citealt{Carr2018}). To counteract this, SALT would need to assume a bi-conical outflow geometry and orient it away from the line of sight. We will explore this scenario further in Section~\ref{sec:discussion}.

At 0.208 Gyr, SALT partially captured the velocity field up to $r = 0.4 R_{\rm vir}$ but failed to reproduce the density field, which exhibits an inflection point and is not well described by a power-law. This discrepancy explains the behavior of the flow rate predictions at this lookback time in Figure~\ref{fig:mpe_radial}. While SALT better captures the radial dependence of $\dot{E}_{\rm out,Si^+}^F$ at small radii—since energy scales with the cube of velocity—it fails to reproduce the radial behavior of $\dot{M}_{\rm out,Si^+}^F$, which depends equally on both the density and velocity fields. This last case underscores the importance of accurately modeling the radial structure of both fields to recover the correct flow rates.

The respective variations on the density and velocity fields, averaged over radius, are 0.2, 0.3 dex, 1.9, 0.03 dex, and 1.2, 0.25 dex for lookback times 0.745, 0.604, and 0.208 Gyr, respectively.

\subsection{Column Density}

\begin{figure}
	\centering
	\includegraphics[width=\columnwidth]{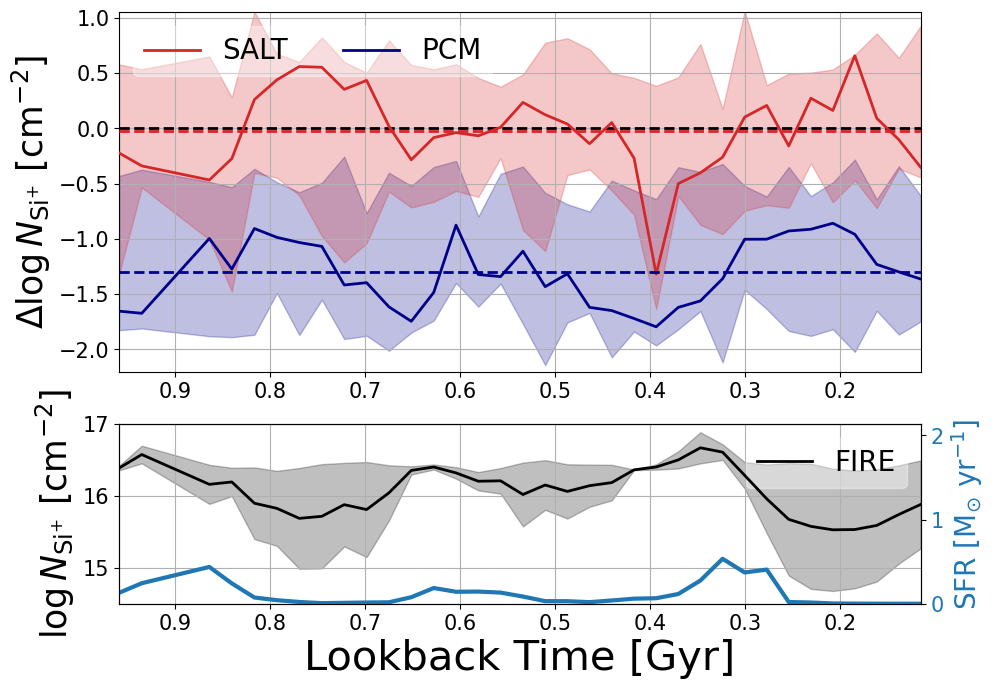}
	\caption{The median LOS SALT (red) and PCM (blue) predictions of the column density. FIRE-2 values were derived through ray tracing, restricting to particles along the chosen LOS. The \textbf{\emph{Top Panel}} shows the median residuals of the logarithm of the column density as a function of lookback time, while the \textbf{\emph{Bottom Panel}} displays the logarithm of the total column density measured in FIRE-2. Error bars represent the absolute deviation from above and below these values. Average values appear as dashed horizontal lines.  The SFR
(light blue) is shown on the right vertical axis.
vertical axis.  The PCM systematically underestimates the column density by 1.3 dex on average, with the discrepancy increasing at higher column densities.  We did not detect a bias in the SALT estimate on average.}
	\label{fig:columnd_density}
\end{figure}

Having identified discrepancies between the SALT and PCM estimates of the flow rates compared to the FIRE-2 values, we now examine other properties that contribute to their calculation. We begin with the column density and assess how well both models recover the FIRE-2 values, as shown in Figure~\ref{fig:columnd_density}. The upper sub-panel presents the median difference in the logarithm of the column densities for each model as a function of lookback time, while the lower sub-panel displays the median FIRE-2 values over time.

The PCM systematically underestimates the column density, with the discrepancy increasing at higher column densities. We measure an average bias of 1.3 dex with an uncertainty of 1.3 dex. These results align with \cite{Jennings2025}, who reported a similar underestimate of 1.25 dex in $N_{Si^+}$ in an analogous study using RAMSES simulations. In contrast, the SALT estimations show no systematic bias, though they still exhibit an uncertainty of 1.3 dex.

\subsection{Geometry}

\begin{figure}
	\centering
	\includegraphics[width=\columnwidth]{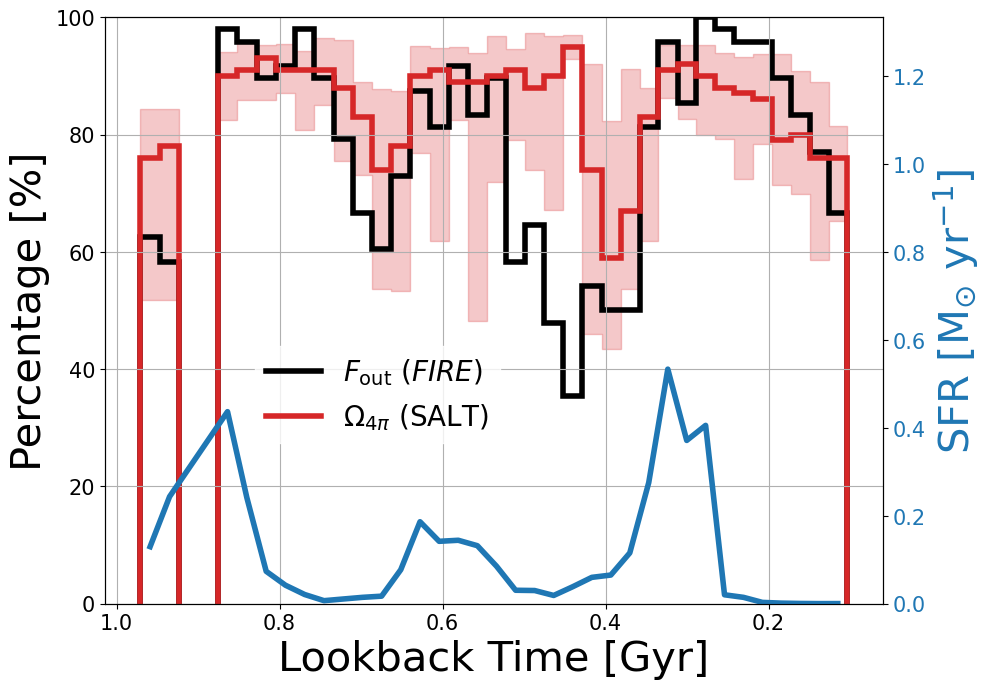}
	\caption{Comparison of the normalized solid angle ($\Omega_{4\pi}$) measured by SALT (red) and the fraction of LOS in FIRE-2 (black) showing outflowing gas in absorption ($F_{\rm out}$), as a function of lookback time. The SALT estimates represent the median across all LOS, with error bars indicating the median deviation above and below these values. The SFR (light blue) is shown on the right vertical axis. Except for the 0.5–0.4 Gyr interval, the two quantities closely trace each other, suggesting that SALT effectively captures the outflow geometry.}
	\label{fig:solid_angle}
\end{figure}

 While the outflow geometry in SALT is well-defined as a radially accelerating bi-conical outflow characterized by an opening angle, $\alpha$, and an orientation angle, $\psi$, the equivalent description in FIRE-2 is less clear. Figure~\ref{fig:geometry} illustrates that FIRE-2 outflows can be highly anisotropic. To assess how well SALT tracks the global outflow geometry, we plot the average solid angle normalized by $4\pi$, $\Omega_{4\pi} = (1-\cos{\alpha})$, alongside the fraction of lines of sight that detect an outflow, $F_{\rm out}$, as a function of lookback time in Figure~\ref{fig:solid_angle}. These values represent medians over multiple lines of sight, with error bars indicating the median deviation above and below the central value.

Overall, we find that $\Omega_{4\pi}$ closely follows $F_{\rm out}$, though there is a significant discrepancy between 0.5 and 0.4 Gyr. As shown in Figure~\ref{fig:geometry}, this period corresponds to the weakest outflow rates, which would be reflected in weak spectral lines—a point we will explore further in Section~\ref{sec:discussion}. Given this, it is unsurprising that SALT does not perform well in this regime.

For further context, the $\Omega_{4\pi}$ values are also listed alongside the corresponding sky maps in Figure~\ref{fig:geometry}. Since these sky maps are projected over radius, they incorporate gas from multiple star formation episodes as well as ISM gas. Consequently, a perfect correlation between $\Omega_{4\pi}$ and $F_{\rm out}$ is not expected. However, the observed relationship provides strong evidence that SALT is able to extract meaningful information about the gas geometry in FIRE-2 from the mock spectral lines.

\section{Discussion}\label{sec:discussion}

In Section~\ref{sec:results}, we tested the ability of the semi-analytical line transfer (SALT) model and partial covering model (PCM) to predict the flow rates and various other properties of an intermediate mass dwarf simulation (m11c) taken from the “core” FIRE-2 suite.  In this section, we dive deeper into the measurements and offer our interpretation of the results.  Given the vast difference in their associated parameter spaces, we discuss the performance of each model separately, focusing on their strengths, weaknesses, and potential for improvement.     

\subsection{PCM Performance}

The PCM shows an average bias of 1.3 dex, consistently underestimating the column density of the outflows within the range  $15 < \log{(N^F_{\rm Si^+}[\rm cm^{-2}])} < 17$, regardless of lookback time, as shown in Figure~\ref{fig:columnd_density}. This result aligns with \cite{Huberty2024}, who found that the PCM systematically underestimated the column density by one to two orders of magnitude over a similar range when testing against idealized turbulent outflows derived from the SALT parameter space (see their Figure 8). They proposed that this error was primarily due to instrumental smoothing, which leads the PCM to measure an apparent optical depth (see \citealt{Savage1991}).

\cite{Jennings2025} also found similar results when testing the PCM against mock spectra derived from RAMSES simulations. They hypothesized that dust attenuation could explain the bias observed in the PCM’s estimate of the column density. Their study was largely influenced by \cite{Mauerhofer2021}, who observed significant dust radiation transfer effects in line profiles drawn from the RAMSES simulations. To explore whether dust contributed to the bias in their PCM predictions, \cite{Jennings2025} compared the dust-attenuated column density ($e^{-\tau_{\rm dust}}N_{\rm Si^+}$) to the PCM’s predictions.  They found a reduction in the bias, but it was nevertheless still present.  They concluded that dust likely plays a role but was not the only factor influencing the bias.  In fact, for a homogeneous medium, normalization should eliminate the influence of dust in unsaturated absorption lines \citep{Carr2021a}. Therefore, we suspect that if dust were the only factor playing a role, then the column density attenuated by dust would be less than the measured value, not greater.

To test the impact of dust on our results, we reran our column density tests with the dust-to-gas ratio set to zero. In the dust-free scenario, the PCM underestimated the FIRE-2 column density by 1.2 dex, or only about 0.1 dex less than before (as shown in Figure~\ref{fig:columnd_density}).  This suggests that dust has only a minimal impact on our study, implying that normalization mitigates the majority of its influence on the line profiles. 

Given the negligible impact of dust on our results and the strong agreement with \cite{Huberty2024}, who studied in a controlled setting, isolating the impact of instrumental smoothing, we conclude that instrumental smoothing accounts for the majority of the bias in the PCM model's prediction of the FIRE-2 column density.   

We measure a roughly uniform dispersion of 1.3 dex in our estimate of the column density within the range $15 < \log{N^F_{\rm Si^+}[\rm{cm}^{-2}]} < 17$.  We attribute this error to sources of uncertainty such as blue emission in-filling, deviations from the Gaussian fitting functions, an inability to correctly distinguish the outflow from ISM absorption, variations in the density field along different lines of sight (see Equation~\ref{eq:Jensen's_Inequality}), and deviations from a spherical outflow geometry (see Figure~\ref{fig:geometry}).  However, the relatively large column densities and fluorescent channels for each of the Si II lines considered in our study, should mitigate the amount of blue emission in-filling.  We propose that this uncertainty more accurately captures the true uncertainty when fitting the PCM to real spectral lines, and propose it as an alternative to the values often derived from constraints on its parameter space.

The PCM provides a reasonable approximation of the flow rates at $r = 0.15\ R_{\rm vir}$ for several snapshots in Figure~\ref{fig:mor_vs_time}, but it systematically overestimates the FIRE-2 values at this and larger radii for most snapshots. We suspect that the agreement at specific radii is coincidental. In fact, if the PCM had accurately estimated the column density, it would have predicted even higher values for $\dot{M}^F_{\rm out,Si^+}$, further amplifying the discrepancy.  For these reasons, we find that the best interpretation of $\dot{M}^P_{\rm out,Si^+}$ is as the maximum flow rate over all radii.  This interpretation also extends to $\dot{P}_{\rm out, Si^+}^P$ and $\dot{E}_{\rm out, Si^+}^P$. 

Figure~\ref{fig:mor_vs_time} shows that the PCM’s flow rate estimates remain relatively constant over time, though they may tend to increase slightly with rising SFR. Given that the PCM only accounts for the absorption component of the line profile, this lack of temporal evolution is unexpected, as absorption profiles in our study varied significantly over time. We suspect this may stem from not converting the apparent optical depth to the true optical depth. \cite{Huberty2024} found that for $\log{N_{\rm Si^+}} > 15$, large variations in the true column density resulted in only minor changes in the recovered apparent column density. This effect arises because the measurement is based on the logarithm of the line profile (see Equation~\ref{eq:AOD}).  In addition, this behavior may reflect the fact that the PCM flow rate's correspond to values integrated over the full extent of the wind.  Lastly, the potential relationship between higher flow rate estimates and increasing SFR may indicate difficulty in distinguishing the outflowing component of the spectrum from the ISM.

In summary, failing to convert the apparent optical depth to the true optical depth in the PCM leads to an underestimate in the column density which can exceed an order of magnitude at column densities $\log{N} [\rm cm^{-2}] > 15$.  This results in an underestimation of flow rates, which we propose to best represent the maximum values with radius.  We conclude that significant improvements could be made by developing a correction to convert the apparent optical depth to the true value.  Although techniques have been developed for this purpose, they typically assume point-like sources and are limited to either low column densities (e.g., $\log{N} [\rm cm^{-2}] < 13$, \citealt{Savage1991}) and/or specific velocity structures (e.g., \citealt{Jenkins1996}).  We are unaware of corrections appropriate for the relatively large column densities associated with the metals that trace the CGM in star-forming galaxies (i.e., $\log{N} [\rm cm^{-2}] > 15$, \citealt{Carr2023,Xu2022a,Huberty2024}), and postpone the development of such a correction to a future project.  


\subsection{SALT Performance}

SALT appears to overestimate $\dot{M}^F_{\rm out,Si^+}$ as a function of lookback time in Figure~\ref{fig:mor_vs_time}, except for a few key regions, between $0.0-0.25$ Gyr, $0.45-0.55$ Gyr, and $0.7-0.8$ Gyr, or roughly where $\dot{M}^F_{\rm out,Si^+}$ follows a relative maxima.  It performs similarly for $\dot{P}^F_{\rm out,Si^+}$ and $\dot{E}^F_{\rm out,Si^+}$, however, these measurements are less accurate from $0.45-0.55$ Gyr.  Likewise, SALT appears to perform its worst when the SFR peaks.  After comparing Figure~\ref{fig:mor_vs_time} to Figure~\ref{fig:solid_angle}, we see that these time periods also correspond to regions following local maxima in the detection rate of outflows in absorption.  Furthermore, after comparing to Figure~\ref{fig:geometry}, we see that during the time periods when SALT overestimates $\dot{M}^F_{\rm out,Si^+}$, the net radial motion along many of the sightlines is dominated by infalling gas, reflecting the occurrence of metal recycling or galactic fountains following the previous starburst.  Together, these results suggest that the SALT model's performance is sensitive to the timing of the baryonic cycle and performs it's best when the outflow rate is high relative to the SFR and inflow rate, resulting in highly blue-shifted absorption wells with large EWs. 

\begin{figure*}
\centering	\includegraphics[width=\textwidth]{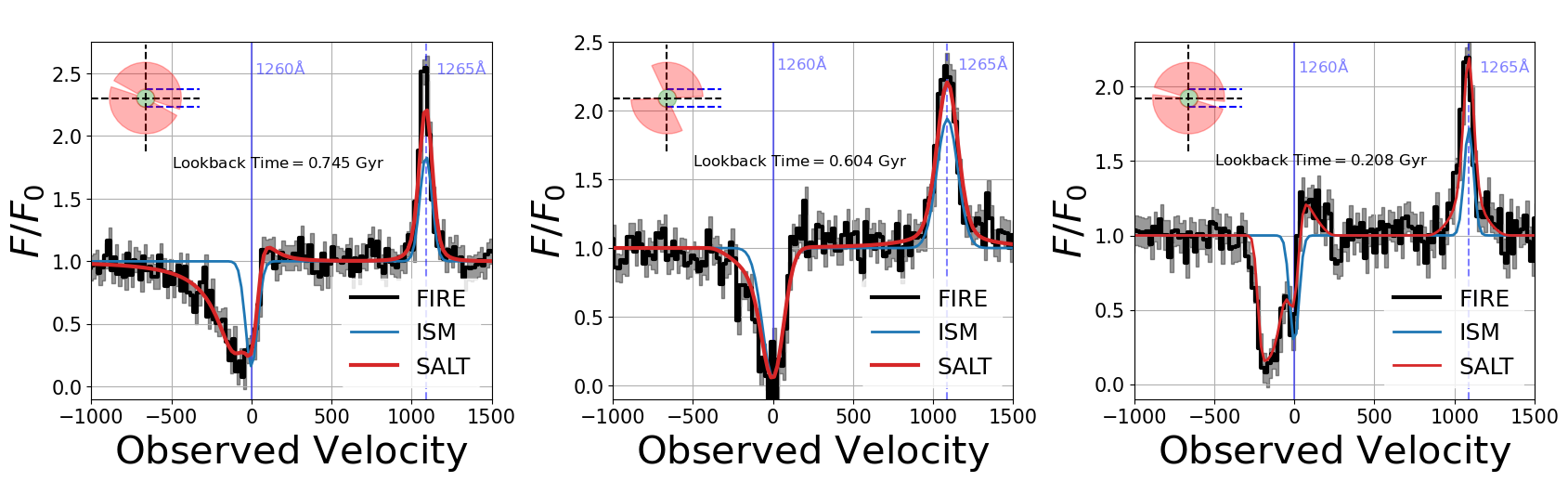}
	\caption{Examples of SALT (red) and ISM (light blue) fits to FIRE-2 (black) spectra at the 1260\AA\ line taken at lookback times 0.745, 0.604, and 0.208 Gyr shown in the \textbf{\emph{Left}}, \textbf{\emph{Middle}}, and \textbf{\emph{Right Panels}}, respectively, and correspond to the rows in Figure~\ref{fig:mpe_radial}.  Emblems representing the best-fit geometry are shown in the upper left corner and are to be viewed from the right.  At lookback time 0.604 Gyr, the SFR reaches a local maximum (see Figure~\ref{fig:mor_vs_time}).  The spectrum is marked by a large ISM component which dominates the absorption and emission spectrum.  Since the ISM feature is removed from the continuum, SALT cannot constrain the outflow to this portion of the spectrum.  Consequently, SALT favors a bi-cone oriented away from the line of sight with high density.  The ISM features are less prominent at lookback times 0.745 and 0.208 Gyr, which show significantly blue-shifted absorption wells.  SALT performs best under these conditions.}
	\label{fig:plot_comparison}
\end{figure*}

To confirm this, we plot representative spectra containing the 1260\AA\ line for each of the snapshots examined in Figures~\ref{fig:mpe_radial} and \ref{fig:density_velocity} in Figure~\ref{fig:plot_comparison} with the left, middle, and right columns correponding to lookback times 0.745, 0.604, and 0.208 Gyr, respectively.  The spectra at 0.745 and 0.208 Gyr show significantly smaller ISM components (light blue) than at 0.604 Gyr. In such cases, SALT better captures the outflow behavior. Conversely, at 0.604 Gyr, SALT struggles because the strong ISM component obscures much of the line profile. In response, SALT favors a bi-conical geometry oriented away from the LOS, minimizing additional absorption beyond the ISM in observed velocity space, leaving other parameters unconstrained. This is consistent with our analysis of the velocity and density fields in Section~\ref{sec:results}.  However, if excess emission is present beyond the ISM contribution, SALT compensates by increasing the bi-cone density. This explains why SALT overestimates flow rates during SFR peaks but performs better when the absorption spectrum is asymmetrically blue-shifted. 

\begin{figure*}
\centering	\includegraphics[width=\textwidth]{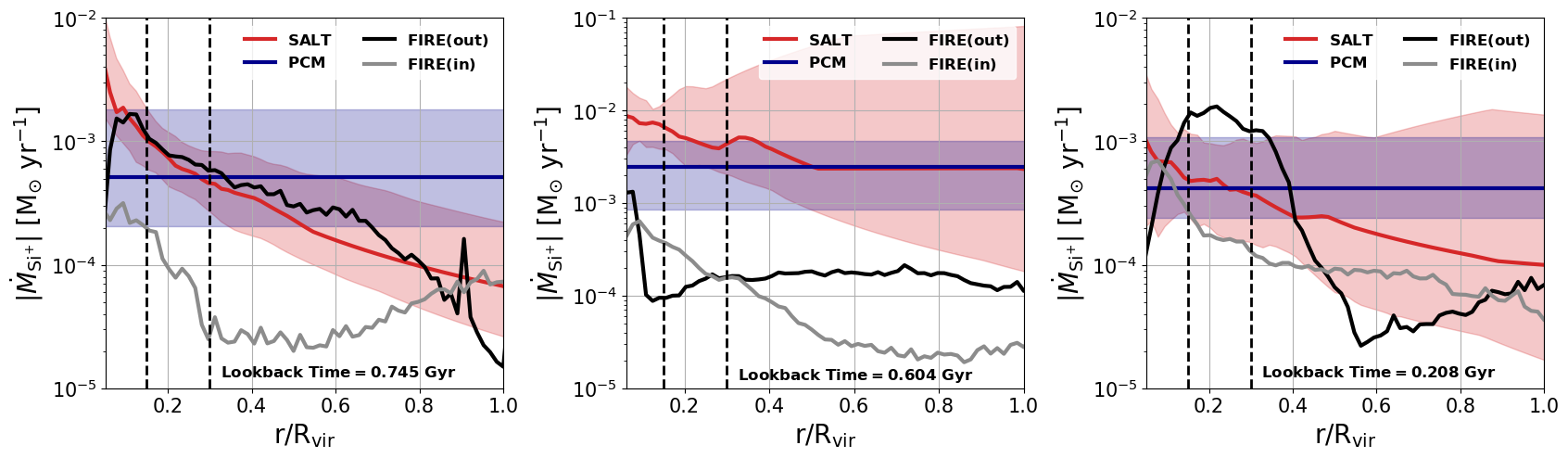}
	\caption{Examples of SALT (red) and ISM (light blue) fits to FIRE-2 (black) spectra at the 1260\AA\ line taken at lookback times 0.745, 0.604, and 0.208 Gyr shown in the \textbf{\emph{Left}}, \textbf{\emph{Middle}}, and \textbf{\emph{Right Panels}}, respectively, and correspond to the rows in Figure~\ref{fig:mpe_radial}.  At lookback time 0.604 Gyr, the SFR reaches a local maximum (see Figure~\ref{fig:mor_vs_time}).  The spectrum is marked by a large ISM component which dominates the absorption and emission spectrum.  Since the ISM feature is removed from the continuum, SALT cannot constrain the outflow to this portion of the spectrum.  Consequently, SALT favors a bi-cone oriented away from the line of sight with high density.  The ISM features are less prominent at lookback times 0.745 and 0.208 Gyr, which show significantly blue-shifted absorption wells.  SALT performs best under these conditions.}
	\label{fig:inflow}
\end{figure*}

We investigate the origin of the ISM component in Figure~\ref{fig:inflow}, which compares the magnitude of the radial inflow rate ($\dot{M}_{\rm in,Si^+}^F$) to that of the outflow rate ($\dot{M}_{\rm out,Si^+}^F$). We suspect that at smaller radii ($r < 0.1 R_{\rm vir}$), $\dot{M}_{\rm in,Si^+}^F$ is influenced by galactic fountains and possibly the ‘breathing modes’ seen in FIRE-2 \citep{El-Badry2016}, while at larger radii ($r \geq 0.1 R{\rm vir}$), it primarily reflects accretion from the IGM. Since the line profile integrates over the full radial extent of the outflow, both sources of inflowing gas can affect SALT’s ability to interpret the profile. Figure~\ref{fig:inflow} shows that SALT performs best when $\dot{M}^F_{\rm out,Si^+}$ is significantly higher than $\dot{M}^F_{\rm in,Si^+}$.

These results suggest that the “ISM” absorption features in Figure~\ref{fig:plot_comparison} arise from a combination of inflowing and outflowing gas. \cite{Carr2022a} proposed that such “hidden” inflows could explain the relatively low detection rate of inflows in observational surveys (e.g., \citealt{Martin2012}) compared to simulations (e.g., \citealt{Angles-Alcazar2017}). This additional scattering in galactic inflows likely causes SALT to overestimate the amount of scattering material or the gas density in outflows. While we used the F-test to discard many such spectra, some remained due to slight deviations from a perfect Gaussian centered on the absorption well. Additionally, since emission probes the entire CGM, biases introduced by enhanced emission in galactic inflows will affect all lines of sight—not just those lacking blue-shifted absorption components \citep{Carr2022a}.

Figures~\ref{fig:mpe_radial} and \ref{fig:density_velocity} suggest that radially averaged quantities can, at least in some cases, be constrained from down-the-barrel spectral lines when there is a net blue-shift in absorption. However, deviations from the simple power-law assumptions in SALT—such as the density field at a lookback time of 0.745 Gyr—can lead to failures in recovering the true underlying radial profile. Expanding the range of profile types to better represent the density field may improve performance. In addition, incorporating more complex velocity fields, like those studied by \cite{Fielding2022} and \cite{Li2024}, could account for a broader range of behaviors. That said, the excellent fit in the right panel of Figure~\ref{fig:plot_comparison}, which corresponds to at least some lines of sight near the median radial quantity in Figure~\ref{fig:density_velocity}, suggests a significant degree of degeneracy.


The absence of a measurable bias in the SALT model’s estimate of $N^{F}_{\rm Si^+}$ in Figure~\ref{fig:columnd_density} is surprising, given that \cite{Carr2023} reported a bias when recovering column densities from turbulent outflows within the SALT parameter space. That study found that turbulent line broadening introduced a particularly strong bias in outflows with column densities $\log{N} [\rm cm^{-2}] > 16$. We suspect that we failed to detect a bias because either our uncertainties in our measurements were so large that we failed to resolve it or the inclusion of an ISM component mitigated the impact of turbulent line broadening. In the latter case, the key difference between spectra generated under the Sobolev approximation and those incorporating turbulence is the presence of excess absorption outside of resonance, which in the SALT model produces a characteristic “tail” at low observed velocities (see \citealt{Carr2023}, Figure 14). The ISM component could effectively mask this feature. Finally, we note that our results are likely sensitive to the chosen launch radius. Figures~\ref{fig:mpe_radial} and \ref{fig:density_velocity} indicate that our adopted star formation radius, $R_{\rm SF}$, defined as the 3D stellar half-mass radius, does not always precisely align with the actual onset of the outflow.       

The bi-conical outflow geometry in SALT appears to constrain the global geometry of the CGM in FIRE-2 in a meaningful way.  As shown in Figure~\ref{fig:solid_angle}, the solid angle of the bi-cone estimated by SALT traces the detection rate of outflows in absorption, which we deem a reasonable measure of the FIRE-2 flow geometry.  However, it is likely that SALT would overestimate the opening angle of the outflow if there is extra emission emanating from galactic inflows.  This could explain the discrepancy in Figure~\ref{fig:solid_angle}, occurring from 0.5-0.4 Gyr.  This is difficult to visualize in Figure~\ref{fig:geometry}, however, as the all sky maps represent net values obtained along a given radius and a single sight line may contain both inflowing and outflow gas.  Nevertheless, we view these results as a strong indication that SALT can constrain the global geometry of the CGM spectral lines observed down-the-barrel (as proposed by \citealt{Carr2018}).  

This could have strong implications, e.g., for constraining the global ionizing escape fraction ($f_{esc}^{LyC}$) from galaxies which is thought to be highly anisotropic \citep{Cen2015}.  \cite{Kimm2014} observed an offset between sSFR and $f_{esc}^{LyC}$ with lookback time in simulations (see their Figure 4), which they connected to the onset of outflows.  This relationship was also verified in simulation m11c by \cite{Ma2020}.  The simple bi-cone describing the flow geometry in SALT appears to be sufficient to describe both the gas distribution and column density of the outflows to potentially verify the relationship reported by \cite{Kimm2014}.          

In summary, we find that the SALT model's ability to constrain the properties of galactic outflows is sensitive to the timing of the baryon cycle.  SALT performs best when the magnitude of the flow rates are high relative to that of the inflow rates and SFR; the inflows contribute to an ISM feature which masks the absorption profile of the outflows.  In observations, this corresponds to line profiles with large net blue shifts in absorption.  The flow rates in FIRE-2 display a wide range of radial behavior.  When this behavior overlaps with the SALT parameter space, we find that the SALT+Bayesian fitting procedure is able to recover it.  However, SALT is also able to achieve reasonable looking fits to spectra which correspond to flows with radically different outflow behavior.  This suggests that there is substantial degeneracy requiring rigorous model fitting and high-resolution, high signal-to-noise data to break.  We did not detect a bias in the SALT estimates of the FIRE-2 column densities which we attribute to the masking of the line profile at low observed velocities by an ISM feature or a high degree of uncertainty in our measurements.  Finally, we find that SALT is able to recover meaningful information about the global geometry of the outflows in FIRE-2 as the solid angle of the bi-cone constrained by SALT traces the detection rate of outflows in absorption.

\subsection{The Impact of Turbulent Line Broadening}

\begin{figure*}
\centering	\includegraphics[width=\textwidth]{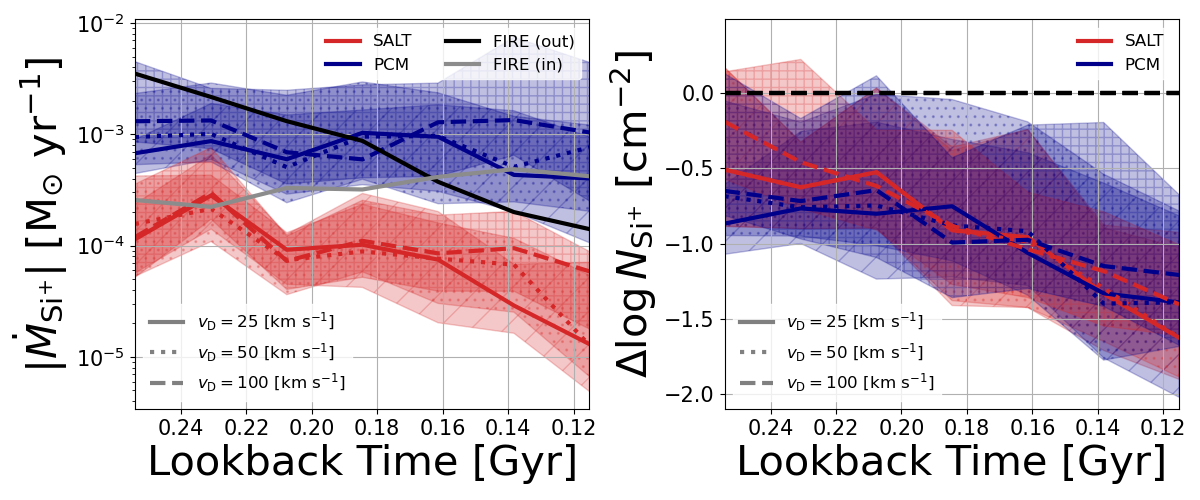}
	\caption{SALT predictions for $\dot{M}_{\rm out,Si^+}^F$ (\textbf{\emph{Left Panel}}) and $N_{\rm Si^+}^F$ (\textbf{\emph{Right Panel}}) for varying values of fixed total Doppler broadening ($v_{\rm b} = 25, 50, 100\ \rm km\ s^{-1}$). The format follows that of Figures~\ref{fig:mor_vs_time} and \ref{fig:columnd_density}. SALT tends to underestimate both $\dot{M}_{\rm out,Si^+}^F$ and $N_{\rm Si^+}^F$ at higher turbulence levels, though there is little variation between the different Doppler broadening parameters. This bias is more pronounced when the outflow rate decreases or when the inflow rate increases. Although the PCM bias has decreased compared to the $v_b = 10\ \rm km\ s^{-1}$ case shown in Figure~\ref{fig:columnd_density}, it remains present.}
	\label{fig:mor_turb_cd}
\end{figure*}

The frequency dependence or width of the interaction cross section between a photon and atom in COLT is set by the Doppler parameter, $v_{\rm b} = v_{\rm th} + v_{\rm turb}$, where $v_{\rm th}$ and $v_{\rm turb}$ are the velocity dispersion due to thermal and turbulent motion, respectively.  While $v_{\rm th}$ can be determined from the gas temperatures reported by FIRE-2 using the usual kinetic theory, $v_{\rm turb}$ must be set by hand.  We chose to set $v_{\rm turb} = 10\ \rm km\ s^{-1}$ to represent the cool gas traced by $\rm Si^+$ \citep{Chen2023}.  It may be informative, however, to investigate how larger values of $v_{\rm b}$ impact our results.  For instance, the $v_{b}$ values associated with higher gas temperatures and higher ionization state lines could easily exceed the parameter ranges probed by our study.  Here we investigate how our results depend on the Doppler width by retesting SALT and the PCM on mock spectra derived from an outflow episode occurring from lookback times $0.11-0.25$ Gyr with a fixed Doppler parameter $v_{b} = 25, 50, 100\ \rm km\ s^{-1}$.


We show the PCM and SALT model predictions of $\dot{M}_{\rm out,Si^+}^F$ and $N_{\rm Si^+}^F$ in the left and right panels of Figure~\ref{fig:mor_turb_cd}, respectively.  In both cases, the SALT model tends to underestimate the FIRE values.  The bias in $N_{\rm Si^+}^F$ appears to increase with decreasing $\dot{M}_{\rm out,Si^+}^F$ or when $|\dot{M}_{\rm in,Si^+}^F|$ increases.  We attribute this behavior to the Sobolev approximation which is more accurate when the outflow speed is high relative to the velocity dispersion (see Equation 24 in \citealt{Carr2023}).  From our discussion in the previous section, one might expect turbulence to result in an overestimate of the column density and flow rates.  However, at these Doppler widths, the entire line profile is so dominated by absorption, there is little emission appearing outside the ISM feature.  Hence, SALT is free to run to the low density end of it's parameter space.    

The performance of the PCM appears to improve when increasing $v_{\rm b}$ from $10\ \rm km\ s^{-1}$ to $25\ \rm km\ s^{-1}$ when comparing Figures~\ref{fig:columnd_density} and ~\ref{fig:mor_turb_cd}. The improvement starts to dwindle, however, when going from $50\ \rm km\ s^{-1}$ to $100\ \rm km\ s^{-1}$.   This behavior likely reflects the measurement of the apparent optical depth, which depends on instrumental smearing, not the underlying velocity structure of the wind \citep{Spitzer1978,Savage1991}.  Thus, while a general smoothing of the line profile due to natural turbulence occurring in the outflows can act to diminish the impact of instrumental smearing, it is insufficient to entirely eliminate its impact in our study.

\subsection{Reconciling the Work of Huberty et al. (2024)}

A large motivation for this work was the study of \cite{Huberty2024} which compared the SALT and the PCM estimates of the outflow properties in the CLASSY galaxies.  In this subsection, we discuss our interpretation of their results in light of our own findings.         

\cite{Huberty2024} showed that the PCM predicted systematically lower column densities than the SALT model, reaching up to two orders of magnitude in the most optically thick lines observed in CLASSY.  The original estimates of the column densities by the PCM covered roughly the same range as our estimates of the FIRE-2 column densities, or between $14<\log{N}\ [\rm cm^{-2}] <16$. Therefore, we suspect that the biases observed in Figure~\ref{fig:columnd_density} apply to the PCM estimates of the outflows in the CLASSY galaxies.  In addition, the true uncertainties in the measurements are likely much larger than those recovered from the PCM parameter space (see Figure~\ref{fig:mor_vs_time}).  In other words, the PCM under fits the data.     

\cite{Huberty2024} noted that the nature of the bias, due to the PCM's measurement of the apparent optical depth, depends on how instrumental smearing changes the shape of the initial line profile. Hence, the scale of the bias depends on the underlying physical description of the outflow.  This was noted as a potential limitation of the tests performed by \cite{Huberty2024} on idealized outflows taken from the SALT parameter space.  While our analysis on the FIRE-2 simulations suffers from this same dilemma, it should provide more reassurance to the existence of the bias since the FIRE-2 simulations provide a more physically realistic and motivated description of the outflows.

While \cite{Huberty2024} reported a bias in SALT’s estimation of column densities for turbulent flows, we did not observe such a bias in the SALT predictions of FIRE-2 column densities. As discussed above, we suspect this discrepancy is due to the presence of large ISM features, which mask the effects of turbulent line broadening at low velocities. While \cite{Huberty2024} employed a similar model when fitting to real data, their tests against idealized models did not include the ISM feature of SALT. Hence, in cases where the CLASSY galaxies show strong ISM components, we suspect that their models would have overestimated the bias.   


\cite{Huberty2024} found that SALT and the PCM estimated similar flow rates for the CLASSY galaxies.  Our results reaffirm their suspicions, however, that the agreement was likely fortuitous.  Indeed, if the PCM under predicts the column density, then it must also underestimate the flow rate.  The highly blue-shifted absorption lines and high outflow detection rate observed in CLASSY create the best conditions for SALT. Therefore, we deem the flow rates presented by both \cite{Huberty2024} and \cite{Xu2022a} to be reasonable.     

It is unclear whether SALT can reliably constrain the radial dependence of the flow rates in CLASSY.  As previously stated, the large net blue shifts in CLASSY create the best conditions for the SALT model to succeed.  As shown in Figure~\ref{fig:density_velocity}, under these conditions, SALT can recover the radial dependence of the density and velocity fields of the FIRE-2 outflows to properly constrain the flow rates.  However, there exist outflows with radial profiles which lie outside the SALT parameter space in FIRE-2.  Albeit, the most extreme of these scenarios occur when the FIRE-2 simulations undergo periods of heavy metal recycling by way of galactic fountains.  This is certainly not the case for the majority of objects in CLASSY.  Therefore, we deem it possible that SALT was able to accurately recover the radial properties of the flow rates in CLASSY, as long as the velocity and density fields are reasonably approximated as simple power-laws.      

While SALT was able to accurately recover the geometry of the CGM in FIRE-2, SALT's performance with the CLASSY galaxies was likely hindered by the observing aperture.  To accurately recover the geometry of an outflow, SALT requires the absorption profile and the extended emission incurred from scattered photons \citep{Carr2018}.  In other words, SALT must “see” the full extent of the outflow to know what it looks like.  Given the small size of the COS aperture, many of the objects in CLASSY are aperture-limited \citep{James2022}.  Thus, we suspect that SALT could only successfully determine the geometry of the most distant objects in CLASSY.

Lastly, due to the more stringent observing conditions of the CLASSY galaxies—such as the limited observing aperture—we suspect that the uncertainties in Figure~\ref{fig:columnd_density} underestimate the true systematic uncertainty in the CLASSY measurements by SALT. The uncertainties reported by \cite{Huberty2024} are likely more appropriate for CLASSY.

\subsection{Caveats}

\begin{figure*}[htbp]
    \centering
    \subfigure{
        \includegraphics[width=\columnwidth]{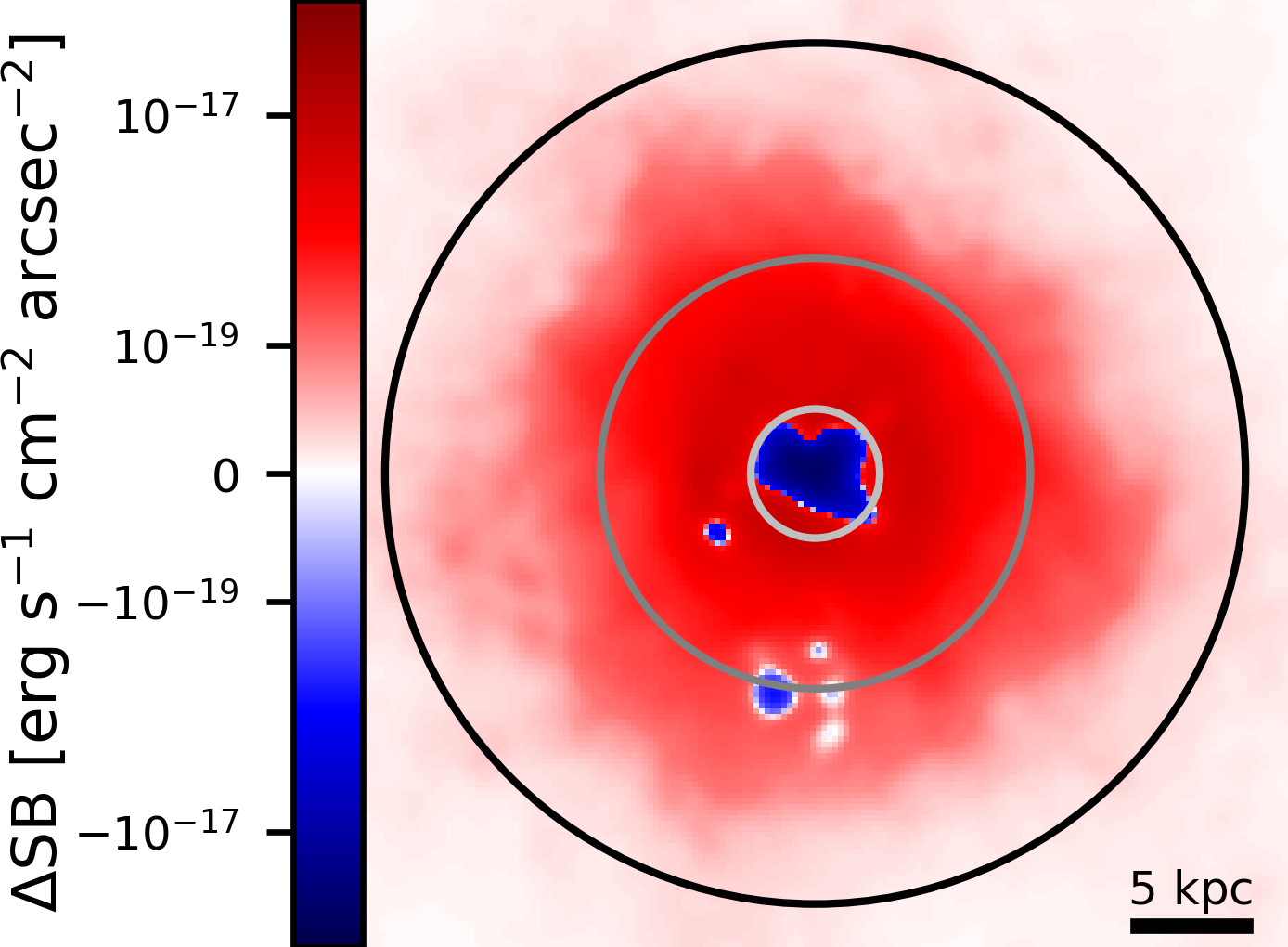}
    }
    \hfill
    \subfigure{
        \includegraphics[width=\columnwidth, height=0.735\columnwidth]{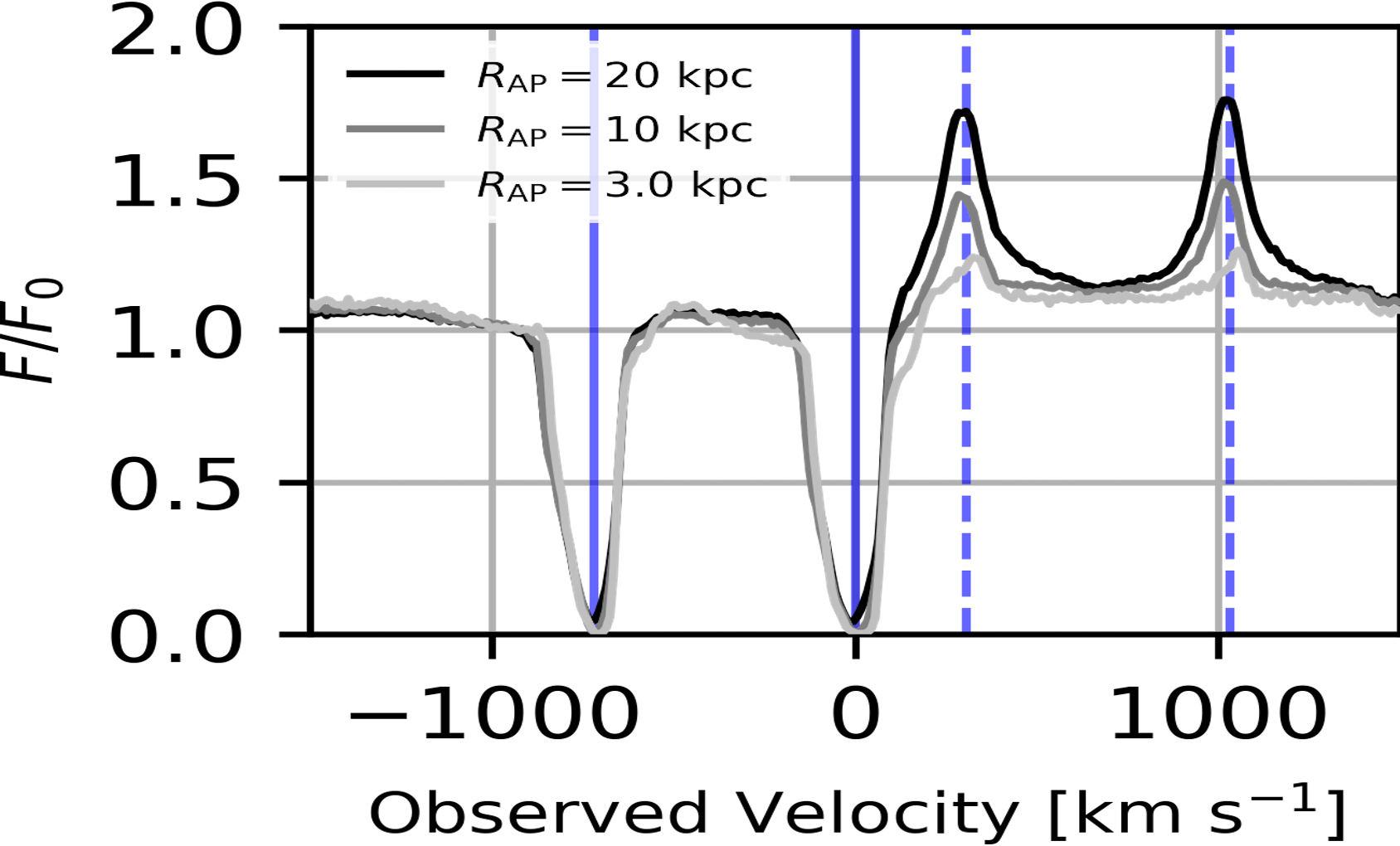}
    }
    \caption{An image of Si II 1190\AA, 1193\AA\ emission obtained from a mock IFU data cube extracted from simulation m11c by COLT.  The \emph{\textbf{Left Panel}} shows the continuum subtracted galaxy image: red regions appear only as emission, while blue regions contribute to both absorption and emission in spectral lines. The rings represent three different aperture sizes and their captured spectra are shown in the \emph{\textbf{Right Panel}}.  By measuring how the emission profile grows with projected radius, one can obtain tighter constraints on the velocity and density fields of outflows.}
    \label{fig:IFU}
\end{figure*}

There are a number of important caveats to consider in this work.  First and foremost, we neglected the ionization structure of the wind by assuming all atoms occupied the same ionization state.  This assumption could potentially lead to radially different structures of the density and velocity fields of the outflows (e.g., see \citealt{Gray2019,Sarkar2022}).  Constraining the velocity and density structure of the winds is especially important for the performance of SALT, whereas the PCM should work independently of the physical structure underlying the wind.  Shocks leading to jump discontinuities in the density field, e.g., would not be accounted for in the power-law structures assumed by SALT and could lead to an error in the estimate of the flow rates at a given radius.  Realistically, accounting for such features in analytical models would be difficult, especially when interpreting down-the-barrel spectra.  Spatially resolved spectra may provide a solution, however.  

By considering the estimation error along multiple lines of sight, we were able to quantify the systematic uncertainty in the measurements of various quantities.  Our mock observations represent an ideal case, however, in which the observing aperture captures the full extent of the CGM and outflows.  In reality, the aperture scale is often a limiting factor (e.g., \citealt{James2022}).  Failing to observe the full extent of the outflow will lead to larger uncertainties on the constraints \citep{Carr2018,Gazagnes2023}.  Furthermore, the uncertainties can be expected to increase with a decrease in resolution and S/N as well.  Therefore, we suggest viewing the systematic uncertainties reported in this paper as lower estimates when fitting to real observations.  

A primary motivation for this work was to explain the apparent discrepancies highlighted by \cite{Huberty2024} in the SALT and PCM estimates of the properties of the galactic outflows in CLASSY.  The FIRE-2 data show several important deviations from the CLASSY sample which are worth reemphasizing. For instance, the outflows in CLASSY achieve much higher outflow velocities than what we observed in FIRE-2.  As noted, these features are ideal for the performance of the SALT model and the validity of the Sobolev approximation.  While the presence of galactic inflows were present in almost every spectrum of the FIRE-2 sample, they are rarely detected in the CLASSY DATA \citep{Huberty2024}.  Thus, while we suspect that the sources of bias identified in this paper persist in the CLASSY data set, the quantitative estimates may not be accurate.  This could be improved in future studies by considering simulations of low redshift galaxies with much higher SFRs.    

\subsection{Future Progress}

There is currently no consensus in the field on the radial structure of outflows.  For example, the SALT model assumes a power law structure for the velocity field, in agreement with the CGOLS simulations studied by \cite{Schneider2020}, but other studies, such as \cite{Chisholm2017b}, assume $\beta$-type velocity laws which are often used in the context of stellar winds \citep{Lamers1999}.  \cite{Fielding2022} and \cite{Li2024} study additional variations, physically motivated by force-balancing arguments. In addition, there are likely substantial deviations from simple analytical forms which scale with the duration of the star formation episode (see Figure~\ref{fig:mpe_radial}, or \citealt{Gray2019}).  Thus, invoking constraints from this study to ensure capturing the true analytical forms of outflows is rather a moot point if it is unclear what the ideal structures are or even if they can be described by simple expressions.   
  
High-resolution Integral Field Unit (IFU) spectroscopy offers a potential solution to this dilemma. By capturing spectra across a spatially resolved 2D field, IFUs enable constraints on the fundamental scale of the outflows (i.e., $R_{\rm SF}$) through absorption, and their spatial extent (i.e., $R_{\rm W}$) in emission.  In Figure~\ref{fig:IFU}, we illustrate the concept with a mock IFU observation derived from a COLT data cube.  We show a continuum subtracted image of Si II 1190\AA, 1193\AA\ emission in the left panel.  Only blue regions contribute to the absorption component of a spectrum, while both the red and blue regions contribute to the emission component.  The various circles represent different aperture sizes.  We show the spectra recovered from each aperture in the right panel.  The smallest aperture, which captures only the absorption region sets the scale of the outflow (i.e., $R_{\rm SF}$) while the largest aperture sets the extent of the outflow.  By modeling how the emission grows with the increasing aperture one can better constrain the density field, velocity field, etc., of the outflow. IFUs also open the door to studying anisotropies which can aid in other applications of line transfer models such as in the study of LyC escape \citep{Chisholm2020}.   

There have already been some exciting results obtained from IFUs regarding outflows.  For example, \cite{Burchett2021} studied a local compact galaxy with KCWI.  Their modeling suggested that the outflow was actually decelerating at large distances from the galaxy, which could not be obtained from the analytical forms assumed in the SALT model.  By decomposing an IFU image into rings, such as those presented in Figure~\ref{fig:IFU}, \cite{Erb2023} was able to simultaneously explain the Lyman alpha profile and high velocity portion of low ionization state absorption lines using the same model --- a problem that had troubled astronomers for several years (see \citealt{Orlitova2018}).  The development of fast radiation transfer codes to study IFU data cubes has already begun (e.g., \citealt{Yuan2023}).  It is important that our instruments follow suite.  The ground based future 30 meter class telescopes, such as the Giant Magellan Telescope (GMT), will lead to major breakthroughs in galaxy formation with optical IFUs.  With a myriad of available absorption lines, UV spectroscopy also offers another amazing opportunity.  As we decide what the next generation UV space telescope will be (e.g., the Habitable Worlds Observatory [HWO]), it is imperative that we consider the dramatic improvements IFUs pose to the field.

\section{Conclusions}\label{sec:conclusions}

In this study, we evaluated the effectiveness of semi-analytical line transfer (SALT) models and empirically based partial covering models (PCMs) in estimating the properties of galactic outflows identified in cosmological zoom-in simulations from the “core” FIRE-2 suite \citep{Hopkins2018} as characterized by \cite{Pandya2021}. \cite{Huberty2024} previously found these models to disagree in their predictions of outflow properties in the CLASSY galaxies \citep{Berg2022}. We fit each model to mock spectra derived in post-processing, assuming photoionization equilibrium conditions with the radiation transfer code COLT \citep{Smith2015} and standard observational procedures. However, we found that outflows could not be identified in most lines of sight in Si II lines, preventing further study. To address this, we combined all atoms into a single ionization state, which greatly enhanced detection rates. We then tested our ability to predict outflow properties in a single intermediate mass dwarf simulation. Specifically, we fit both SALT and PCM to the resonant and fluorescent lines of the Si II 1190Å, 1193Å, 1260Å, 1304Å, and 1527Å multiplet at a spectral resolution of $20\rm \ km\ s^{-1}$ and $S/N = 10$ in the continuum, to resolve the discrepancies observed by \cite{Huberty2024}. Here are our conclusions.

\begin{itemize}
    \item The PCM consistently underestimates the column densities of the FIRE-2 simulations, which range between $15 < \log{N}[\rm cm^{-2}]<17$, by an average of 1.3 dex. We found that these results were only marginally impacted by the dust content of the simulations.  These findings reaffirm \cite{Huberty2024}'s tests, suggesting that the bias can be attributed to the apparent optical depth.  Additionally, our results align with \cite{delaCruz2021}, who observed similar behavior in optically thick lines. We also note an average dispersion of 1.3 dex in the measurements.
    
    \item The PCM model measures an outflow rate similar to the FIRE-2 values at a radius of $r = 0.15R_{\rm vir}$, with an uncertainty of 0.55 dex. Given that the flow rates depend directly on the underestimated column density, this agreement appears fortuitous. If the PCM had accurately recovered the column density, the predicted flow rates would be greater.  Hence, we propose viewing the flow rate estimates of the PCM as representing the maximum value over all radii. The differing definitions of the flow rates (cf. Equations~\ref{eq:PCM_MOR}-\ref{eq:PCM_EOR} to Equations~\ref{eq:SALT_MOR}-\ref{eq:SALT_EOR}) explain how they could derive similar flow rates for the CLASSY galaxies while obtaining vastly different column densities. 
    
    \item The median estimate of the flow rates by the PCM shows little variation with the LOS. We attribute this to two reasons: first, the PCM measures an apparent optical depth, which becomes insensitive at large column densities ($\log{N}\ [\rm cm^{-2}] > 15$); second, in the definitions of the flow rates (Equations~\ref{eq:PCM_MOR}-\ref{eq:PCM_EOR}), the calculations correspond to integrated values, drawn from the full extent of the wind.

    \item We found that SALT accurately predicts the FIRE-2 column densities within an uncertainty of 1.3 dex as a function of lookback time. The absence of a bias, which is commonly observed when estimating column densities in turbulent outflows using the Sobolev approximation, may be attributed to either the masking of the low-velocity portion of the spectrum by ISM absorption or the inability to resolve the bias due to the large uncertainties in our measurements.  

    \item SALT requires a net blue shift in the absorption profile to accurate measure the FIRE-2 flow rates.  In the FIRE simulations, this occurs when the outflow rate is much greater than the inflow rate.  When successful, we find that SALT accurately measures the mass outflow rate ($\dot{M}^F_{\rm out,Si^+}$), the momentum outflow rate ($\dot{P}^F_{\rm out,Si^+}$), and the energy outflow rate ($\dot{E}^F_{\rm out,Si^+}$) to 0.36(0.63), 0.56(0.56), and 0.97(0.80) dex at $0.15(0.30)R_{\rm vir}$, respectively.  

    \item Under optimal conditions of strong outflows, reflected by large blue-shifted asymmetries in absorption lines, we find that SALT can recover the radial dependence of the velocity and density field of the outflows in FIRE-2.  When this occurs, SALT is also able to capture the radial behavior of the flow rates.  There are instances, however, where the radial behavior of the outflows in FIRE-2 are not well described by simple power-laws, and lie outside the SALT formalism.  When this occurs, SALT can fail to capture the behavior of all three quantities.          
    
    \item We visualized the time evolution of the outflow geometry by plotting stereographic projections of the CGM as a function of lookback time.  We find that the outflows start off filling relatively small solid angles (in some cases resembling bi-cones) which monotonically increase to form spheres in parallel with the increasing mass outflow rate.  The solid angle estimated by SALT traces the temporal evolution of this geometry which corresponds to the photon escape fraction in hydro-simulations, which is out of phase with the SFR \citep{Kimm2014,Ma2020}.  This result reassures our ability to infer global CGM properties, including the ionizing radiation escape fraction, from combined absorption and emission line diagnostics.    
    
\end{itemize}

\begin{acknowledgments}
We acknowledge Alaina Henry, Michael Jennings, Tim Heckman, Chris Howk, and Harley Katz for insightful discussions and Simon Gazagnes for providing information on the CLASSY galaxies.  A.S. is supported by NASA grants HST-AR-17559.015 and HST-AR-17859.010.  R.C. acknowledges in part financial support from the start-up funding of Zhejiang University and Zhejiang provincial top level research support program. C.C. acknowledges the Center for Computational Astrophysics (CCA) at the Flatiron Institute for hosting him for a portion of his time spent on this project through their (virtual) Pre-Doctoral Program.  C.C. is supported by NSFC grant W2433001 and the NSFC Talent-Introduction Program.  The Scientific Computing Center (SCC) at the Flatiron Institute provided the majority of computational resources used in this work.  We also acknowledge Nick Carriero of the SCC for providing general computing advice which improved the efficiency of our codes.  Additional work was also performed using the resources of the Minnesota Supercomputing Institute (MSI) and the SilkRiver Supercomputer of Zhejiang University.  
\end{acknowledgments}

\bibliography{bibliography}{}
\bibliographystyle{aasjournal}



\end{document}